\documentclass{article}

\usepackage{microtype}
\usepackage{graphicx}
\usepackage{subcaption}
\usepackage{enumitem}
\usepackage{subfloat}
\usepackage{array}
\usepackage{booktabs} 
\usepackage{xspace}
\usepackage{hyperref}

\usepackage{booktabs,multirow,tabularx}
\usepackage[table]{xcolor}

\usepackage[preprint]{icml2026}

\usepackage{amsmath}
\usepackage{amssymb}
\usepackage{mathtools}
\usepackage{amsthm}
\usepackage{xcolor}

\usepackage[capitalize,noabbrev]{cleveref}

\theoremstyle{plain}

\theoremstyle{definition}

\theoremstyle{remark}

\newcommand{\icon}{\raisebox{-7pt}{\includegraphics[width=0.048\textwidth]{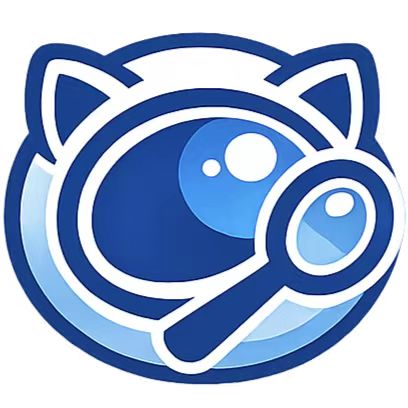}}}
\newcommand{\name}[0]{\textsc{MarkCleaner}\xspace}
\usepackage[textsize=tiny]{todonotes}

\definecolor{bgcolor}{RGB}{56, 90, 70}
\icmltitlerunning{MarkCleaner: High-Fidelity Watermark Removal via Imperceptible Micro-Geometric Perturbation}

\begin{document}

\twocolumn[
  \icmltitle{\icon~\textcolor{bgcolor}{\name}: High-Fidelity Watermark Removal via \\ Imperceptible Micro-Geometric Perturbation  
 }



  \icmlsetsymbol{equal}{*}
  \icmlsetsymbol{cor}{\dag}
  \begin{icmlauthorlist}
    \icmlauthor{Xiaoxi Kong}{szu}
    \icmlauthor{Jieyu Yuan}{nku}
    \icmlauthor{Pengdi Chen}{szu}
    \icmlauthor{Yuanlin Zhang}{nku}
    \icmlauthor{Chongyi Li}{nku}
    \icmlauthor{Bin Li}{szu}

  \end{icmlauthorlist}

  \icmlaffiliation{nku}{VCIP, CS, Nankai University}
  \icmlaffiliation{szu}{Guangdong Provincial Key Laboratory of Intelligent Information Processing, Shenzhen University}

  \icmlcorrespondingauthor{Jieyu Yuan}{jieyuyuan.cn@gmail.com}
  \icmlcorrespondingauthor{Bin Li}{libin@szu.edu.cn}

  \icmlkeywords{Machine Learning, ICML}

  \vskip 0.3in
  \renewcommand\twocolumn[1][]{#1}%
\vspace{-6mm}%
\begin{center}
    \centering
	\includegraphics[width=1\textwidth]{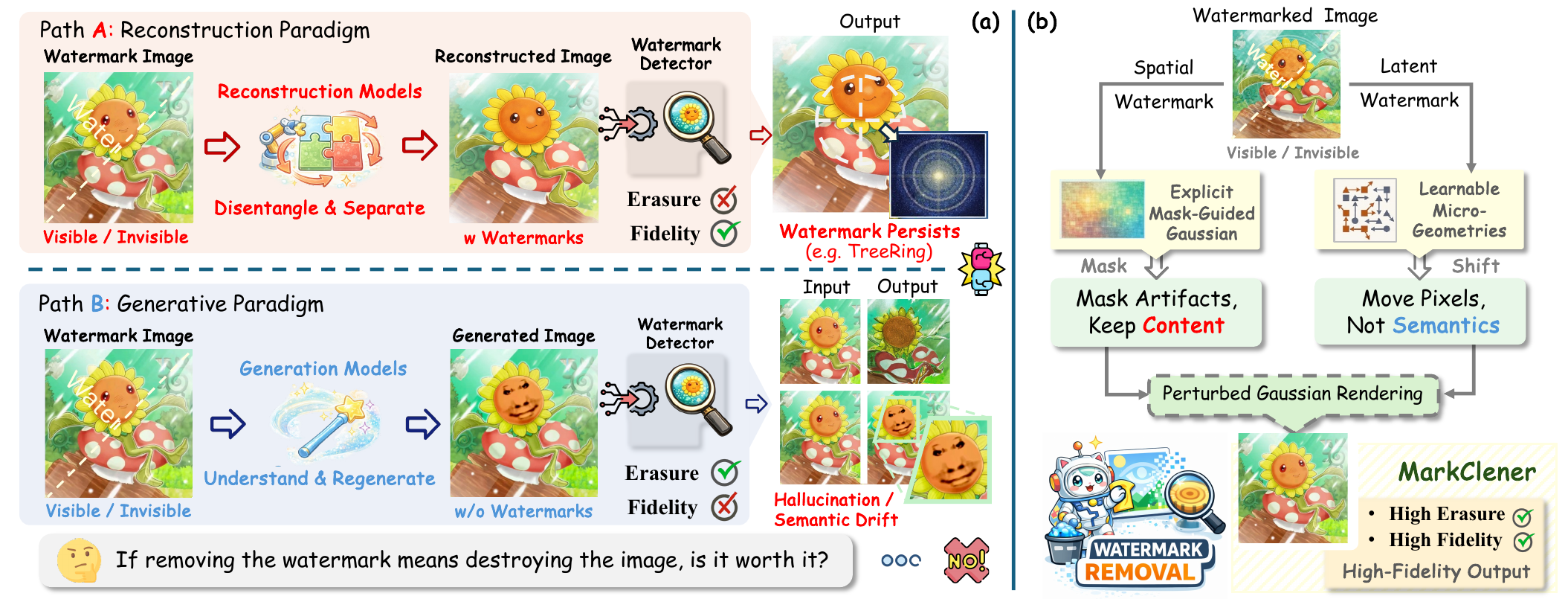}
	\vspace{-5mm}
	\captionof{figure}
    {\textbf{MarkCleaner remove watermark without compromising visual fidelity.} 
    Existing paradigms face a fundamental dilemma: high fidelity or high erasure, but not both. (a) Trade-off between fidelity and erasure in existing methods: Reconstruction-based methods (\textit{Path A}) have high fidelity but fail to remove the semantic watermark, while generative methods (\textit{Path B}) successfully erase watermarks but introduce severe semantic drift, hallucinating content, or altering details. (b) Our MarkCleaner offers a unified solution for both visible and invisible watermarks by resolving this issue via mask-guided encoding and geometric perturbation, enabling clean removal without content distortion.}
    \label{fig:teaser}
    \vspace{3mm}%
\end{center}%

]



\printAffiliationsAndNotice{}  

\begin{abstract}
Semantic watermarks exhibit strong robustness against conventional image-space attacks. In this work, we show that such robustness does not survive under micro-geometric perturbations: spatial displacements can remove watermarks by breaking the phase alignment. Motivated by this observation, we introduce MarkCleaner, a watermark removal framework that avoids semantic drift caused by regeneration-based watermark removal. Specifically, MarkCleaner is trained with micro-geometry-perturbed supervision, which encourages the model to separate semantic content from strict spatial alignment and enables robust reconstruction under subtle geometric displacements. The framework adopts a mask-guided encoder that learns explicit spatial representations and a 2D Gaussian Splatting–based decoder that explicitly parameterizes geometric perturbations while preserving semantic content. Extensive experiments demonstrate that MarkCleaner achieves superior performance in both watermark removal effectiveness and visual fidelity, while enabling efficient real-time inference. \textit{Our code will be made available upon acceptance.}

\end{abstract}

\section{Introduction}

Digital watermarking is a cornerstone technology for image copyright attribution and content traceability. With the rapid advancement of generative AI~\cite{LDM_2022_CVPR, gpt_achiam2023, labs2025flux}, the boundary between human-created and AI-generated content has become increasingly blurred, intensifying the need for reliable watermarking. Recent watermarking technology has shifted from visible overlays to invisible watermarks~\cite{review_2024,gowal2025synthid}. 
Among invisible watermarks, recent generative methods embed signals by intervening in the sampling process or modulating the initial latent distribution. Methods such as TreeRing~\cite{Treerings_nips_2023} are often classified as \textbf{semantic watermarks}~\cite{semantic_forgery_CVPR_2025}, as they embed watermark information through integration with the high-level semantic structure of generated images rather than through direct manipulation in pixel or frequency domains.
Such deep coupling significantly increases the difficulty of removal, making semantic watermarks a particularly challenging benchmark for evaluating the limits of watermark purification methods. To evaluate their practical reliability, it is essential to explore potential removal mechanisms.

From the perspective of technical implementation, existing watermark removal methods can be categorized into two paradigms, both facing inherent limitations against semantic watermarks, as illustrated in Figure~\ref{fig:teaser}(a). 
\textbf{Reconstruction-based removal methods} attempt to eliminate watermarks by reconstructing clean images through denoising and compression techniques that prioritize pixel-level fidelity, effectively suppressing the embedded watermark signal~\cite{BM3D_TIP_2007, compressai_ARXIV_2020, sadre_www_2025}.
However, semantic watermarks are embedded by modifying the initial latent noise, making them inherently entangled with latent structural representations. 
As a result, higher reconstruction fidelity preserves more watermark information, directly contradicting the objective of watermark removal.
\textbf{Generation-based methods} leverage generative models to regenerate image content, effectively disrupting the original latent structure and thereby erasing semantic watermarks well~\cite{DIFF_2024_NIPS, CtrlRegen_iclr_2025, NFPA_nips_2025}. 
The generative processes are typically computationally expensive due to their iterative nature, requiring substantial resources and time for high-quality results.
Most importantly, the inherent stochasticity of the denoising process inevitably induces hallucinations, semantic drift, and loss of fine-grained details. 
This reveals a fundamental dilemma in existing approaches: \textit{high fidelity or high erasure, but not both}. If removing the watermark entails destroying the image content, is such a trade-off truly acceptable?

In this work, we found that semantic watermarks are fundamentally sensitive to geometric perturbation. 
This key insight originates from a frequency-domain perspective. 
It is well established that the phase spectrum predominantly encodes structural information, while the magnitude spectrum captures textural attributes~\cite{phase_IEEE_1981,fda_CVPR_2020,fourier_cvpr_2021}. 
TreeRing watermark is embedded by modulating the spectrum within the Fourier transformed latent space, causing watermark signals to be tightly coupled with the image structure and to remain stable under signal processing attacks. 
However, geometric transformations manifest as phase shifts in the frequency domain.
Even minimal geometric perturbations are sufficient to disrupt the precise phase alignment, while remaining virtually imperceptible to human observers. 
Therefore, watermark removal can be reframed from content regeneration to geometric perturbation, enabling high-fidelity removal without compromising visual content consistency.

Building upon this insight, we propose \textbf{MarkCleaner}, a novel watermark removal framework, as shown in Figure~\ref{fig:teaser}(b). The core principle is to train the network to learn micro-geometric perturbations while maintaining visual consistency, such that the output is visually consistent with the input yet geometrically displaced to remove the watermark. 
Specifically, we introduce a mask-guided encoding strategy operating at frequency and spatial levels to suppress watermark patterns while extracting semantic representations. For decoding, we propose a novel 2D Gaussian Splatting (2DGS) renderer that explicitly parameterizes micro-geometric perturbations to break spatial alignment. During training, the model is supervised using geometrically perturbed targets rather than original inputs, guiding the decoder to apply these transformations while the encoder learns perturbation-invariant semantic features. To maintain semantic consistency, we incorporate self-supervised visual feature alignment as an additional constraint.

\textbf{In summary, our contributions are as follows:}
	\vspace{-8pt}
\begin{itemize}[leftmargin=*]\setlength\itemsep{-0em}
    \item \textbf{Uncovering the Geometric Vulnerability of Semantic Watermarks.}
    We uncover an overlooked phase sensitivity in semantic watermarks, demonstrating that even imperceptible micro-geometric perturbations can disrupt watermark detection. This finding reveals a fundamental geometric vulnerability and motivates a paradigm shift from content regeneration to geometry-based watermark removal.
     \item \textbf{A Novel Watermark Removal Framework.} We propose MarkCleaner, which combines mask-guided encoding with geometry-perturbed supervision to train models that maintain semantic consistency under spatial displacement. Our framework achieves SOTA performance across 12 diverse watermarking schemes without requiring prior knowledge of watermark types or detection mechanisms.
    \item \textbf{Perturbed 2D Gaussian Rendering.} We introduce a 2DGS-based decoder that adopts differentiable rasterization to learn spatial displacements, enabling real-time inference with pixel-level structural preservation.
    \vspace{-6pt}
\end{itemize}

\section{Related Work}

\textbf{Image Watermarking.}
 Image watermarking enables the verification of provenance by embedding visible and imperceptible signals. 
 Existing invisible watermarks can be categorized by when the watermark is embedded and where it resides in the representation. \textit{\textbf{Post-processing Watermarks}} operate on generated images after synthesis. Traditional methods~\cite{DwtDct_2007,DWT-DCT-SVD_2008} embed watermarks in the frequency domain using wavelet and discrete cosine transforms. Deep learning-based methods \cite{RivaGAN_2019,SSL_icassp_2022,StegaStamp_CVPR_2020} adopt an encoder-decoder structure. Recent works~\cite{WOFA_cvpr_2025,VINE_iclr_2025} include progressive training and leveraging generative priors for robustness against partial image theft and AI-based editing.
\textit{\textbf{In-processing Watermarks}} integrate watermarking into generation, creating intrinsic semantic coupling. Model finetune method ~\cite{StableSignature_ICCV_2023} fine-tunes VAE decoders, while latent manipulation approaches~\cite{Treerings_nips_2023,RingID_eccv_2025,semantic_iccv_2025,GaussianShading_CVPR_2024,t2smark_NIPS_2025} structure diffusion noise patterns in Fourier transformed space or preserve the statistical properties of Gaussian noise. This coupling provides superior robustness to compression and noise.

\textbf{Watermark Removal Attacks.}
Existing removal strategies can be categorized from an implementation perspective into three groups \cite{CtrlRegen_iclr_2025}. Signal processing approaches apply transformations such as JPEG compression and filtering~\cite{waves_iclr_2024}, but primarily target pixel-level perturbations with limited effectiveness against semantic watermarks. Regenerative methods leverage diffusion models~\cite{CtrlRegen_iclr_2025,NFPA_nips_2025} or VAEs~\cite{VAEattack_CVPR_2020} to reconstruct images, yet suppressing semantic watermarks often requires substantial noise injection, risking semantic drift. Adversarial attacks optimize perturbations to mislead detectors~\cite{sok_arxiv_2024}, but typically require white-box access~\cite{leveraging_iclr_2024} or surrogate training~\cite{tune_cvpr_2023,diffuction-attack_iclr_2024}, limiting practical applicability.
These paradigms focus on modifying image content rather than spatial configuration. Our analysis suggests that semantic watermarks may be sensitive to geometric perturbations, motivating watermark suppression through minimal geometric transformations while preserving semantic fidelity.

\textbf{Gaussian Splatting.} 
Gaussian Splatting (GS)~\cite{3DGS_tog_2023} represents scenes using explicit Gaussian primitives, enabling real-time rendering with high visual fidelity. Recent works extend GS to the 2D image domain, modeling images as continuous mixtures of 2D Gaussians~\cite{Gaussiansr_aaai_2025}. 
Benefiting from its high effectiveness, image-based 2DGS has been rapidly adapted for image compression~\cite{GaussianSR_eccv_2024}, super-resolution~\cite{GSASR_iccv_2025}, and image inpainting~\cite{2dgsinpainting_2025}, demonstrating advantages in continuous spatial modeling and efficient rendering. Moreover, Gaussian-based representations exhibit strong robustness and flexibility under geometric transformations~\cite{DeblurringGS_eccv_2024, ContinuousSR_2025}. 
We consider that the precise position parameterization of GS provides a mechanism for modeling geometric perturbations, making it well-suited for disrupting rigid watermark patterns while preserving visual coherence. Motivated by these properties, we explore the potential of 2DGS for watermark removal.

\section{Preliminary Analysis}
\label{sec:analysis}

In this section, we present a theoretical and empirical analysis of the geometric vulnerability in semantic watermark using TreeRing~\cite{Treerings_nips_2023} as an example.

\subsection{Watermark Formalism and Verification}
Consider a latent diffusion model (LDM) generating images $I \in \mathbb{R}^{H \times W \times 3}$ from noise $z_T \in \mathbb{R}^{h \times w \times c}$ via DDIM sampling~\cite{DDIM_ICLR_2021} and VAE decoding. TreeRing embeds a signature pattern $k^* \in \mathbb{C}^{|M|}$ into the Fourier coefficients of $z_T$ within a low-frequency mask $M$. Specifically, watermarked noise $\tilde{z}_T$ satisfies $\mathcal{F}(\tilde{z}_T)[M] = k^*[M]$ and $\mathcal{F}(\tilde{z}_T)[\bar{M}] \sim \mathcal{N}(0,1)$ where $\bar{M}$ denotes the complement. The watermarked image $I_w$ is then generated via denoising and VAE decoder. 

Detection recovers the initial noise latent $\hat{z}_T$ from a suspected image $I$ via VAE encoding and DDIM inversion. The watermark presence is verified by the normalized $L_1$ distance between the recovered Fourier coefficients and the $k^*$ over the frequency mask $M$:
\begin{equation}
d = \frac{1}{|M|} \sum_{\mathbf{u} \in M} \big|k^*_{\mathbf{u}} - \hat{Z}_{\mathbf{u}}\big|, 
\label{eq:detection}
\end{equation}
where $\hat{Z} = \mathcal{F}(\hat{z}_T)$, $\mathbf{u}$ represents the frequency coordinates within the latent Fourier space, and $d < \tau$ indicates a successful detection for a predefined threshold $\tau$. This mechanism relies on the alignment between $k^*$ and the recovered $\hat{Z}$ within $M$.

\subsection{The Geometric Vulnerability of Latent Phase}
\begin{figure}[t]
    \centering
    \subfloat[Clean]{
    \includegraphics[width=0.31\columnwidth]{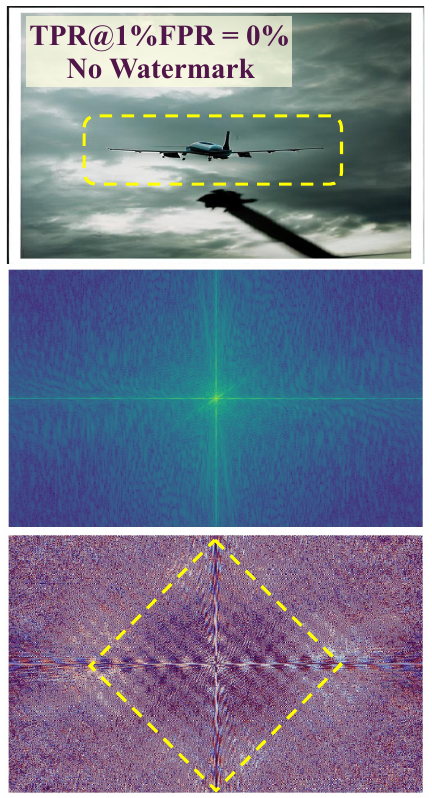}}   
    \label{fig:phase_a}
    \subfloat[Watermarked]{
    \label{fig:phase_b}
    \includegraphics[width=0.31\columnwidth]{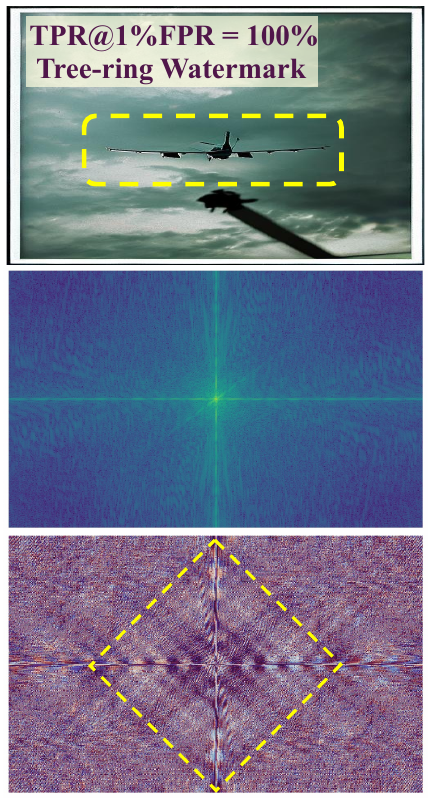}}
    \subfloat[Transformed]{
    \label{fig:phase_c}
    \includegraphics[width=0.31\columnwidth]{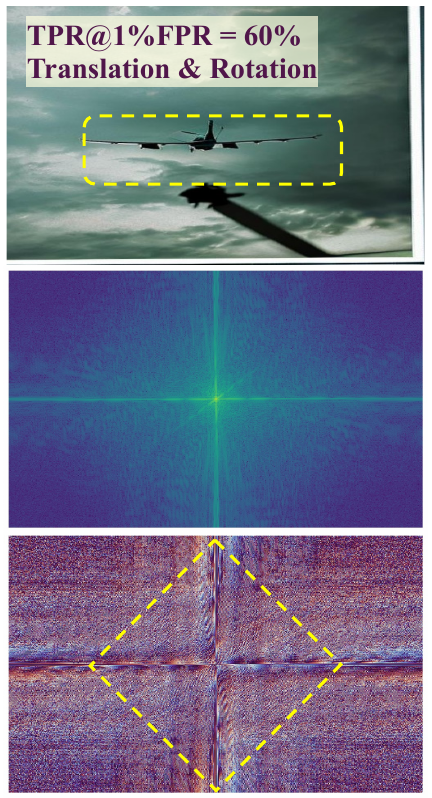}}
    \caption{\textbf{Frequency-domain analysis of phase perturbation caused by geometry.} Starting from the same initial latent, we show (a) standard generation, (b) generation with latent watermark, and (c) geometric transformation of (b). Each column displays the RGB image with latent amplitude and phase spectra. Geometric transformation disrupts watermark-induced phase ripples while preserving amplitude structure, indicating watermark invalidation arises from phase modulation rather than content alteration. TPR@1\%FPR computed over 1000 images. (Zoom in for details.)
    }
    \label{fig:phase_visualization}
\end{figure}

The strided architecture of VAE creates a critical vulnerability: pixel-domain transformations $\mathcal{T}_{\boldsymbol{\Delta}, \theta}$ trigger a cascade of phase-domain scrambling~\cite{Aliasfree_2025_CVPR}.
A translation $\boldsymbol{\Delta}$ in pixel space induces a fractional shift $\boldsymbol{\delta} = \boldsymbol{\Delta}/d$ in the latent space ($d=8$). According to the Fourier Shift Theorem, this spatial displacement propagates as a phase modulation:
\begin{equation}
\hat{Z}'_{\mathbf{u}} = \hat{Z}_{\mathbf{u}} \cdot e^{i \Phi(\mathbf{u}, \mathcal{T})},
\end{equation}
where $\Phi(\mathbf{u}, \mathcal{T}) = -2\pi \big\langle \mathbf{u}, \frac{\boldsymbol{\Delta}}{W} \big\rangle + \theta \cdot \arg(\mathbf{u})$, encoding the combined phase shift, and $\arg(\mathbf{u}) = \operatorname{atan2}(u_y, u_x)$ denotes the angle of frequency vector $\mathbf{u}$.
Although geometric transformations preserve the amplitude spectrum, they systematically rotate each Fourier coefficient in the complex plane. Crucially, the detection metric in Eq.~\eqref{eq:detection} computes the $L_1$ distance between complex values, which expands with phase disagreement: $|k^*_{\mathbf{u}} - \hat{Z}'_{\mathbf{u}}| \approx 2|k^*_{\mathbf{u}}| |\sin(\Phi_{\mathbf{u}}/2)|$. As displacement increases, the phase ramp $\Phi_{\mathbf{u}}$ rotates the latent vectors away from the real-axis manifold of $k^*$, would inflate the $L_1$ error.
Meanwhile, rotation by angle $\theta$ induces spectral rotation, misaligning recovered 
coefficients with the embedded key $k^*_{\mathbf{u}}$.
The resulting destructive interference causes detection score $d$ to drift from $0$ toward mean magnitude $\bar{A} \gg \tau$, triggering detection failure (details in Appendix~\ref{sec:Geometry_analysis}), as empirically validated in Figure~\ref{fig:phase_visualization}:
The amplitude spectra remain nearly identical between watermarked and transformed 
images, confirming preserved visual fidelity, while phase patterns are scrambled 
after 7px translation and $5^{\circ}$ rotation. In the representative example shown, a transformation increases the detection distance. Evaluated across 1000 generated images, this minimal perturbation reduces detection performance from TPR@1\%FPR=100\% to 
60\%, demonstrating that micro-geometric transformations can effectively 
invalidate TreeRing watermark detection while maintaining perceptual quality.

Our analysis reveals that semantic watermarks rely on stable spatial correspondence between the image and latent space. This exposes a critical vulnerability: their detectability can be invalidated through micro-geometric perturbations while preserving perceptual fidelity, suggesting fundamental limitations in spatially locked watermarking mechanisms.
Crucially, this sensitivity is not unique to semantic watermarks. Traditional schemes in both the spatial and frequency domains have long been vulnerable to rotation, scaling, and translation attacks that disrupt embedded signals \cite{Lin_tip_2005}. 
Despite decades of research, recent benchmarks confirm geometric attacks remain a persistent challenge across modern paradigms \cite{waves_iclr_2024}, highlighting spatial alignment as a fundamental rather than a scheme-specific limitation.
Motivated by this observation, we propose MarkCleaner, a framework that exploits this dependency through controlled geometric transformations to achieve effective and unified watermark removal without semantic drift.

\begin{figure*}[t]
	\centering
	\includegraphics[width=0.96\linewidth]{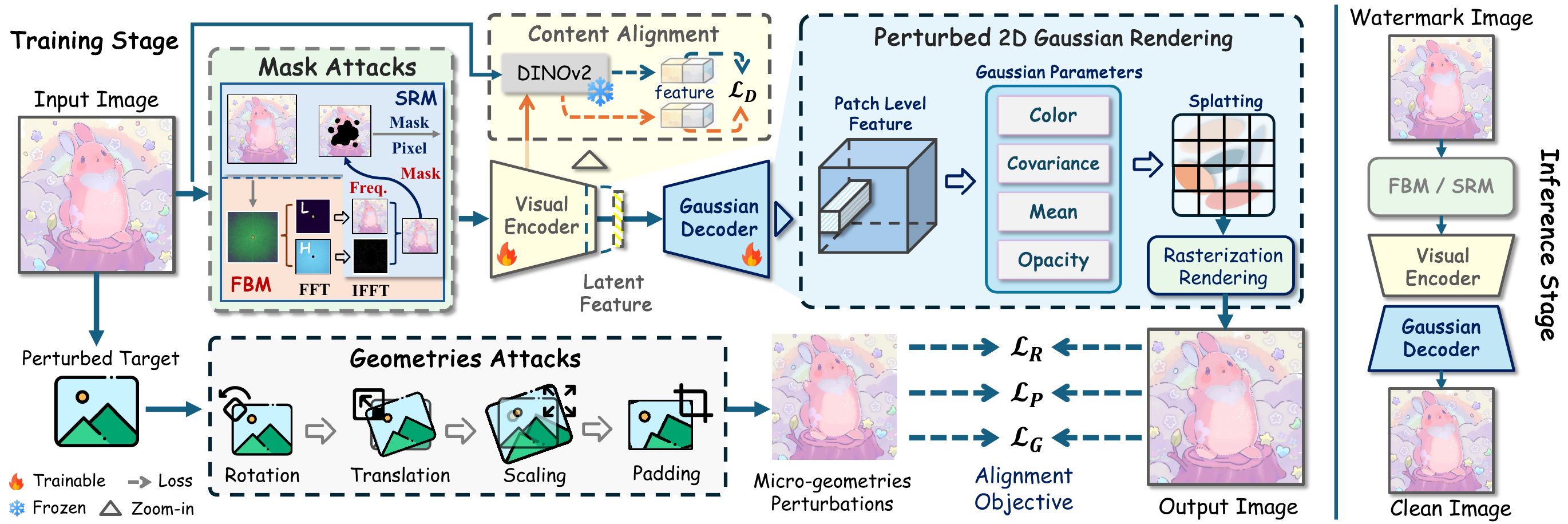}
	\caption{
    \textbf{Overview of MarkCleaner.} \textit{\textbf{Training Stage}}: Input image is first processed by dual-domain masking, including frequency band masking (FBM) and spatial random masking (SRM), then encoded and decoded into per-patch 2D Gaussian parameters (mean, covariance, color, opacity) for differentiable rasterization. The model is supervised using geometrically perturbed targets rather than the original input, with losses $\mathcal{L}_R$, $\mathcal{L}_P$, $\mathcal{L}_G$, and DINOv2-based alignment $\mathcal{L}_D$. \textit{\textbf{Inference Stage}}: Watermarked images pass through the same pipeline, with optional user-provided masks for visible watermark regions, producing semantically consistent yet geometrically perturbed outputs.
    }
	\label{fig:pipeline}
\end{figure*}

\section{Method}
Motivated by the analysis in Sec.~\ref{sec:analysis}, we propose MarkCleaner, a unified watermark removal framework via controlled micro-geometric perturbations. As shown in Figure~\ref{fig:pipeline}, MarkCleaner represents the input image as a set of 2D Gaussian primitives and performs watermark removal by learning imperceptible geometric displacements through differentiable Gaussian rendering.
The framework adopts a UNet-based encoder-decoder architecture. Given an input image, we apply dual-domain mask attacks in both spatial and frequency domains to disrupt potential watermark patterns (Sec.~\ref{sec:mask_encoding}). A visual encoder then extracts multi-scale features, which are decoded into per-patch 2D Gaussian parameters and rendered via differentiable rasterization (Sec.~\ref{sec:2dgs}). During training, the model is not supervised to reconstruct the input. Instead, the rendered output is matched to a micro-geometrically perturbed version of the input, enabling the model to learn geometric displacement while preserving visual content (Sec.~\ref{sec:geo_desync}). To ensure semantic consistency under spatial displacement, we further incorporate feature-level alignment and perceptual constraints (Sec.~\ref{sec:training}).  We describe each component in detail below.

\subsection{Mask-Guided Encoding}
\label{sec:mask_encoding}
To disrupt watermark embedded in the spatial or frequency domains, we apply dual-domain masking before encoding. By randomly masking parts of the input, the network is encouraged to reconstruct the underlying clean content based on global semantic structure.

\textbf{Frequency Band Masking (FBM).} We first transform the input image $I$ into the frequency domain via 2D Fast Fourier Transform (FFT). A circular mask with radius $r$ centered at the spectrum origin is used to separate low- and high-frequency components. During training, we randomly apply either low-pass or high-pass filtering with equal probability. This stochastic frequency masking disrupts spectral watermark signatures (e.g., the ring-shaped patterns in TreeRing) regardless of their embedding location, while preserving sufficient structural information for reconstruction.

\textbf{Spatial Random Masking (SRM).} We then apply spatial masks to disrupt the spatial continuity for clean semantic inpainting. Specifically, we randomly mask $r\%$ of pixels, where both the mask ratio $r$, spatial locations, and region sizes are randomly sampled during training. 
This irregular spatial masking trains the encoder to extract coherent semantic features from incomplete spatial information, which is essential for maintaining content consistency under the geometric perturbations applied during decoding while preventing overfitting to complete image structures. As an additional benefit, users can optionally provide binary masks for visible watermark regions to guide localized removal.

During training, FBM and SRM are composed as stochastic masking operations. When frequency masking is sampled, it is applied before spatial masking. The masked image is still transformed back to the spatial domain before encoding, ensuring consistent input representations between training and inference~\cite{MFM_iclr_2023}.

\subsection{Perturbed 2D Gaussian Rendering}
\label{sec:2dgs}

General pixel-wise decoders generate outputs on a fixed spatial grid, where pixel values are tightly coupled to predetermined locations, limiting flexibility for spatial reorganization \cite{LIIF_cvpr_2021, GaussianSR_eccv_2024}. Due to this rigidity, even a minor sub-pixel shift causes significant semantic misalignment across the entire grid, forcing the decoder to re-synthesize global pixel values to simulate displacement. In contrast, 2DGS represents images as continuous fields of overlapping Gaussians, where each primitive has learnable explicit parameters. 
Crucially, this property decouples geometry  (position, covariance) from appearance(color, opacity). 
The explicit representation is particularly beneficial for watermark removal, since watermark invalidation during evaluation is primarily due to spatial misalignment rather than semantic degradation.
We can learn a micro-offset field applied to Gaussian centers instead of forcing the network to regress complex pixel-level changes to hide watermarks.
Therefore, we introduce 2DGS as our rendering decoder. 

\textbf{Gaussian Representation.} Each 2D Gaussian primitive is characterized by mean $\mu_i \in \mathbb{R}^2$ (spatial position), covariance $\Sigma_i \in \mathbb{R}^{2\times 2}$ (spatial extent), color $c_i \in \mathbb{R}^3$, and opacity $\alpha_i \in \mathbb{R}$. The covariance is parameterized as $\Sigma_i = L_i L_i^\top$ with lower triangular $L_i$ to ensure positive semi-definiteness. The pixel value at position $p$ is computed by summing Gaussian contributions:
\begin{equation}
    I_p = \sum_{i} c_i \alpha_i \exp\left(-\frac{1}{2}(p-\mu_i)^\top \Sigma_i^{-1}(p-\mu_i)\right).
\end{equation}

\textbf{Patch-Level Gaussian Decoding.} The masked image is first encoded by a UNet-based encoder into multi-scale latent features. A Gaussian decoder then predicts per-patch Gaussian parameters from these features. To handle high-resolution images efficiently, we adopt patch-level rasterization by dividing the image into non-overlapping patches, each assigned a dedicated set of Gaussians \cite{2dgsinpainting_2025}. Adjacent patches share overlapping borders that are blended to ensure smooth transitions. The final output is obtained via differentiable rasterization, enabling end-to-end training. Further details are provided in Appendix~\ref{sec:supp_gs}.

\subsection{Geometric Deviation Strategy}
\label{sec:geo_desync}

We apply micro-geometric perturbations to generate a displaced supervision target, enabling the model to learn geometric displacement during training. 

\textbf{Perturbation Pipeline.}
Given an input image $I$, we sequentially apply four geometric transformations to obtain the perturbed target $I_{pert}$:
\begin{itemize}[leftmargin=*]\setlength\itemsep{-0em}
\vspace{-8pt}
    \item \textbf{Rotation}: Slight rotation around the image center.
    \item \textbf{Translation}: Small spatial shift along both axes.
    \item \textbf{Scaling}: Minor scale adjustment followed by resizing to the original resolution.
    \item \textbf{Padding}: Missing regions caused by the above transformations are filled with corresponding content from the original image.
\vspace{-6pt}
\vspace{-8pt}
\end{itemize}
The perturbation magnitudes are intentionally kept minimal (e.g., within a few degrees for rotation and several pixels for translation), ensuring that the induced displacement remains imperceptible to human observers while being sufficient to disrupt the precise spatial and phase alignment exploited by both frequency-based and semantic watermark detectors. 
The network learns to absorb geometric displacement into its generation process by supervising the model to reconstruct $I_{\text{pert}}$ instead of $I$. At inference, this learned behavior naturally produces outputs that are semantically consistent with the input yet geometrically displaced, thereby invalidating watermark decoding without altering visual content.

\textbf{Micro-Geometric Perturbation Supervision.}
Unlike conventional image reconstruction that minimizes the distance to the original input, our training supervises the rendered output $I_{out}$ against the perturbed target $I_{pert}$. This design forces the model to internalize spatial displacements as part of its learned representation, rather than treating them as errors to be corrected. At inference, this learned behavior naturally produces outputs that are semantically consistent with the input yet geometrically displaced, thereby invalidating watermark decoding without altering visual content.

\subsection{Content Alignment.}
\label{sec:training}

\textbf{Self-Supervised Feature Alignment.}
To preserve content and semantic consistency under geometric displacement, we incorporate self-supervised visual features from a frozen DINOv2 encoder~\cite{DINOV2_tmlr_2024} as content anchors. DINOv2 features capture high-level scene structure and exhibit strong invariance to appearance and local perturbations, making them well-suited for our setting.
Specifically, we extract features from the masked input image using the frozen DINOv2 encoder. To accommodate varying mask ratios and spatial corruption, a lightweight MLP is employed to adapt the features before conditioning the Gaussian decoder via Adaptive Layer Normalization (AdaLN)\cite{analn_2019_cvpr}. This conditioning encourages the decoder to preserve global semantic structure while allowing local geometric displacement. In addition, we adopt a DINO-based feature alignment loss based on cosine similarity between the rendered output and the perturbed target, further enforcing semantic consistency under spatial shifts. Details are provided in Appendix~\ref{sec:supp_dino}.

\textbf{Loss Function.}
The total training objective combines multiple complementary losses:
\begin{equation}
    \mathcal{L}_{total} = \lambda_1 \mathcal{L}_{rec} + \lambda_2 \mathcal{L}_{P} + \lambda_3 \mathcal{L}_{G} + \lambda_4 \mathcal{L}_{D},
\end{equation}
where $\mathcal{L}_{rec}$ is the reconstruction loss combining $L_1$ and frequency-domain terms to ensure pixel-level and spectral fidelity, $\mathcal{L}_{P}$ is the LPIPS perceptual loss for high-level visual quality, $\mathcal{L}_{G}$ is the adversarial loss to encourage realistic outputs, and $\mathcal{L}_{D}$ is the content feature alignment loss for semantic consistency. Loss formulations are detailed in Appendix~\ref{sec:supp_loss}.

\begin{table*}[ht]
  \centering
  \caption{Quantitative comparison of robustness against various removal attacks. Results are shown as TPR@1\%FPR / ACC. The last column summarizes the overall performance via mTPR / m$\Delta$A (mean TPR and mean $|\mathrm{ACC} - 0.5|$). Lower values indicate more effective removal, making detection closer to random guessing. Ours achieves the best overall performance across the evaluated removal attacks.}
  \label{tab:all_results}
  \renewcommand{\arraystretch}{1.35}
  \footnotesize %
  \resizebox{1\linewidth}{!}{%
  \begin{tabular}{l|*{6}{c}|*{2}{c}|*{4}{c}|c}
        \hline
    \multirow{2}{*}{\textbf{Attack Type}} & 
    \multirow{2}{*}{\textbf{DwtDct}} & 
    \multirow{2}{*}{\textbf{SSL}} & 
    \textbf{Stega} & \textbf{Stable} & 
    \multirow{2}{*}{\textbf{VINE}} & 
    \multirow{2}{*}{\textbf{WOFA}} & 
    \textbf{Gaussian} & \textbf{T2S} & \textbf{Tree} & 
    \multirow{2}{*}{\textbf{RingID}} & 
    \multirow{2}{*}{\textbf{HSTR}} & 
    \multirow{2}{*}{\textbf{HSQR}} &
    \textbf{mTPR}($\downarrow$) \\
    
    & & & \textbf{Stamp} & \textbf{Sign.} & & & \textbf{Shading} & \textbf{Mark} & \textbf{Ring} & & & &  \textbf{m$\Delta$A}($\downarrow$)\\
    \hline   
    \rowcolor{blue!5} 
    None & .800/.888 & 1.0/1.0 & 1.0/.999 & 1.0/.993 & 1.0/.999 & .977/.832 & 1.0/1.0 & 1.0/1.0 & .943/.970 & 1.0/1.0 & 1.0/1.0 & 1.0/1.0 & .977 / .473 \\
    \hline 
    JPEG Compression & .003/.490 & .243/.680 & 1.0/.997 & .782/.695 & 1.0/.989 & .040/.458 & 1.0/1.0 & 1.0/.998 & .083/.945 & 1.0/1.0 & .980/.988 & 1.0/1.0 & .678 / .353 \\   
    \rowcolor{blue!5} 
    Crop \& Scale & .007/.521 & .927/.875 & .033/.554 & .999/.978 & .013/.505 & .800/.631 & .003/.501 & .000/.501 & .000/.842 & .010/.582 & .070/.597 & .153/.702 & .251 / .154 \\
    Gaussian Blur & .357/.682 & .987/.981 & 1.0/1.0 & .818/.817 & 1.0/.998 & .693/.707 & 1.0/1.0 & 1.0/1.0 & .593/.965 & 1.0/1.0 & 1.0/1.0 & 1.0/1.0 & .871 / .429 \\   
    \rowcolor{blue!5} 
    Gaussian Noise & .000/.505 & .010/.508 & .987/.872 & .000/.543 & 1.0/.901 & .069/.439 & 1.0/.991 & .987/.951 & .000/.925 & .993/1.0 & .103/.578 & .987/.990 & .511 / .267 \\
    Color Jitter & .061/.523 & .517/.769 & .927/.868 & .895/.848 & .953/.884 & .278/.470 & .990/.993 & .953/.973 & .247/.962 & .960/.998 & .927/.967 & 1.0/1.0 &233 / .138\\   
    \rowcolor{blue!5} 
    Rotation & .000/.522 & .973/.918 & .000/.510 & .998/.813 & .010/.497 & .987/.845 & .007/.539 & .000/.499 & .000/.850 & 1.0/1.0 & .103/.578 & .060/.613 & .251 / .154 \\
    Translation & .003/.501 & .980/.923 & .255/.587 & 1.0/.991 & .013/.504 & .822/.963 & .027/.568 & .000/.501 & .000/.888 & .013/.484 & .087/.585 & .057/.563 & .271 / .172\\ \rowcolor{blue!5} 
    \hline
    VAE-B (ICLR, \citeyear{VAE_B_iclr_2018})  & .000/.499 & .760/.818 & 1.0/.999 & .730/.680 & 1.0/.990 & .157/.439 & 1.0/1.0 & 1.0/1.0 & .190/.957 & 1.0/1.0 & 1.0/1.0 & 1.0/1.0 & .686 / .365 \\  
    VAE-C (CVPR, \citeyear{VAEattack_CVPR_2020}) & .000/.496 & .410/.720 & 1.0/.998 & .582/.652 & .997/.959 & .173/.440 & 1.0/1.0 & 1.0/1.0 & .123/.947 & 1.0/1.0 & .990/.993 & 1.0/1.0 & .606 / .351 \\ \rowcolor{blue!5} 
    DA (ICLR, \citeyear{diffuction-attack_iclr_2024}) & .000/.494 & .003/.510 & .043/.531 & .625/.463 & .450/.609 & .477/.479 & 1.0/.987 & .993/.939 & .000/.880 & .927/.996 & .800/.828 & 1.0/.998 & .526 / .226 \\   
    \hline
    CtrlRegen+ (ICLR, \citeyear{CtrlRegen_iclr_2025}) & .003/.494 & .030/.557 & .197/.603 & .023/.479 & .833/.671 & .110/.416 & 1.0/.999 & 1.0/.987 & .001/.885 & .980/.999 & .437/.890 & 1.0/.998 & .468 / .248 \\ \rowcolor{blue!5} 
    UnMarker (S\&P, \citeyear{UnMarker_ISSP_2025}) & .010/.538 & .973/.918 & .990/.946 & .999/.981 & .007/.502 & .997/.917 & 1.0/1.0 & .017/.502 & .010/.948 & .460/.939 & .097/.693 & .033/.575 & .466 / .288 \\
    IRA (CVPR, \citeyear{semantic_forgery_CVPR_2025})& .000/.487 & .680/.759 & .990/.905 & .625/.508 & 1.0/.924 & .100/.418 & 1.0/1.0 & 1.0/1.0 & .800/.975 & 1.0/1.0 & .990/.995 & 1.0/1.0 & .682 / .331 \\ \rowcolor{blue!5} 
    NFPA (NeurIPS, \citeyear{NFPA_nips_2025}) & .000/.500 & .030/.565 & .013/.481 & .043/.494 & .013/.502 & .087/.387 & .003/.514 & .000/.499 & .003/.913 & .030/.672 & .153/.645 & .367/.795 & .061 / .154 \\
    \hline
    \textbf{MarkCleaner (Ours)} & \textbf{.003/.500} & \textbf{.000/.504} & \textbf{.010/.488} & \textbf{.000/.446} & \textbf{.063/.554} & \textbf{.002/.494} & \textbf{.003/.554} & \textbf{.003/.500} & \textbf{.001/.795} & \textbf{.003/.802} & \textbf{.051/.630} & \textbf{.022/.766} & \textbf{.014 / .094} \\
    \bottomrule
  \end{tabular}}
  \newcolumntype{C}{>{\centering\arraybackslash}X}
\end{table*}

\section{Experiments}
\subsection{Experiment Settings}

\textbf{Datasets and Model Setup.} We train MarkCleaner on 5,000 images from MS-COCO-2017~\cite{coco_eccv_2014} and evaluate on 1,000 paired watermarked and clean images from DiffusionDB-2M~\cite{diffusiondb_AMACL_wang2023}. 
All images are generated by Stable Diffusion v2-1~\cite{LDM_2022_CVPR} at $512\times512$ resolution with classifier-free guidance 7.5 and 50 DDIM steps~\cite{DDIM_ICLR_2021}.

\textbf{Watermarking Methods.}
MarkCleaner is evaluated on fifteen publicly released watermarks, including both post-processing (DwtDct~\cite{DwtDct_2007},  SSL~\cite{SSL_icassp_2022}, StegaStamp~\cite{StegaStamp_CVPR_2020}, VINE~\cite{VINE_iclr_2025}, WOFA~\cite{WOFA_cvpr_2025} and in-processing (Stable Sig.~\cite{StableSignature_ICCV_2023}, TreeRing~\cite{Treerings_nips_2023}, RingID~\cite{RingID_eccv_2025}, HSTR~\cite{semantic_iccv_2025}, HSQR~\cite{semantic_iccv_2025}, Gaussian Shading~\cite{GaussianShading_CVPR_2024}, T2SMark~\cite{t2smark_NIPS_2025}) methods. Each baseline explicitly targets robustness, yielding a stringent benchmark for removal performance.

\textbf{Watermark Removal Methods.}
We compare MarkCleaner against 14 representative watermark removal approaches spanning three categories: distortion-based attacks, regeneration-based attacks leveraging generative priors~\cite{VAEattack_CVPR_2020,diffuction-attack_iclr_2024,CtrlRegen_iclr_2025}, and optimization-based attacks~\cite{UnMarker_ISSP_2025}.
All evaluations are conducted in a strictly black-box setting.
To ensure fair comparison, we adopt an iso-quality protocol that aligns output visual fidelity across methods, isolating their relative effectiveness in watermark erasure.
Detailed configurations are provided in Appendix~\ref{Attack_settings}.

\textbf{Implementation and Metrics.}
All models are trained using Adam with a learning rate of $2\times10^{-4}$ and batch size $16$. The patch size is $16\times16$, and each Gaussian is parameterized with a 12-dimensional embedding. Experiments are conducted on four NVIDIA L40 GPUs.
Watermark robustness is evaluated by TPR@1\%FPR (lower is better) and Bit Accuracy (ACC, 0.5 indicates randomization), following prior work~\cite{Treerings_nips_2023,CtrlRegen_iclr_2025}. Image quality is evaluated using FID~\cite{FID_nips_2017}, CLIP score~\cite{clipscore-2021-emnlp}, and LPIPS~\cite{lpips_cvpr_2018}.

\begin{figure*}[!th]
	\centering
	\includegraphics[width=1\linewidth]{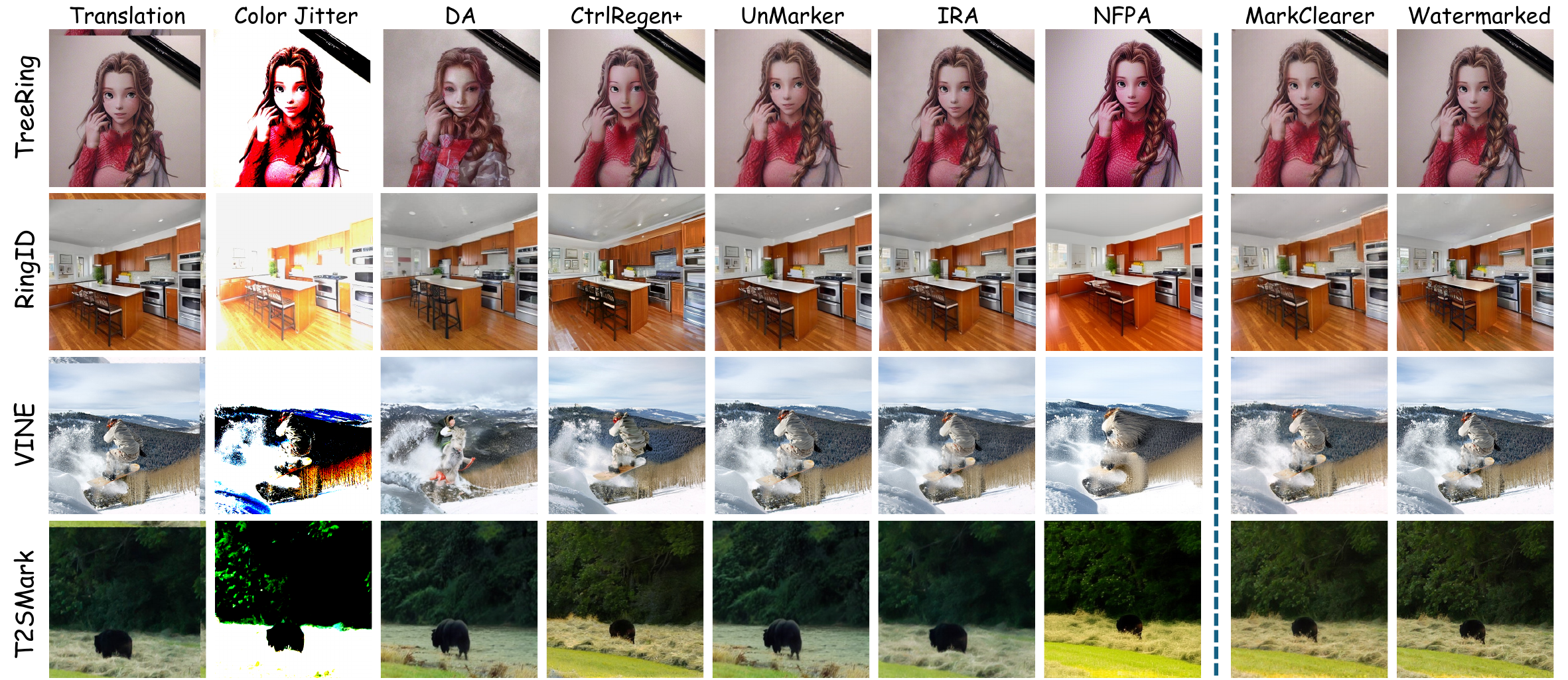}
	\caption{
    \textbf{Qualitative comparison of watermark removal methods.} Traditional pixel-space distortions severely degrade visual quality, while generation-based methods tend to remove watermarks at the cost of semantic drift. Our MarkCleaner achieves more effective watermark suppression with better visual fidelity. More results are provided in the Appendix \ref{sec:sup_visual_results}. 
    }
	\label{fig:result}
    \vspace{-8pt}
\end{figure*}

\subsection{Experiment Results}
We evaluate removal effectiveness, visual quality, and inference speed under an iso-quality protocol where all methods are calibrated to produce comparable image fidelity.

\textbf{Watermark Erasure Performance.}. Table~\ref{tab:all_results} reports watermark removal results of 14 attack methods on DiffusionDB-2M Prompts \cite{SDPrompts}. MarkCleaner consistently achieves the lowest $\text{mTPR}$ and $\text{m}\Delta\text{A}$ across all evaluated categories, demonstrating superior and stable erasure performance. Notably, our method effectively suppresses detection across both semantic watermarks and non-semantic schemes, validating its universality across diverse watermarking paradigms.
We attribute this broad generalization to the displacement mechanism of 2DGS rendering. Traditional detectors rely on rigid pixel-grid alignment for signal extraction. The continuous rasterization in 2DGS, however, introduces localized sub-pixel shifts. These micro-geometric perturbations effectively break the spatial synchronization required by watermark verification, rendering patterns unreadable without targeting specific structures.
In contrast, baseline methods show significant limitations. Distortion-based attacks succeed on fragile watermarks but fail on robust semantic schemes, often retaining near-perfect detection accuracy. Regeneration-based methods improve robustness but remain inconsistent, with detection rates deviating significantly from chance. Optimization-based approaches demonstrate stronger generalization yet still exhibit noticeable residual detectability. MarkCleaner's near-uniform suppression of $\text{TPR}$ and $\text{ACC}$ across all watermark types indicates effective decoupling of artifacts from semantics, achieving reliable universal erasure

\textbf{Removal Fidelity.}
Figure~\ref{fig:result} shows qualitative comparisons of different watermark removal strategies.
Traditional image-space distortions can invalidate semantic watermarks but require aggressive perturbations that severely degrade visual quality.
Generation-based methods can invalidate watermark detection but introduce noticeable semantic drift, such as altered facial attributes (Row~1) or blurred and over-smoothed textures (Row~3).
Moreover, these methods introduce strong color casts and unintended geometric shifts, reflecting the lack of explicit spatial control during regeneration.
In contrast, MarkCleaner removes both visible and invisible watermarks while preserving fine-grained image structure and semantic content.
The outputs remain visually indistinguishable from the original images, without introducing color distortion or perceptible geometric artifacts.
More results are provided in the Appendix~\ref{sec:sup_visual_results}.

\textbf{Inference Efficiency.} Table~\ref{tab:efficiency} compares inference efficiency and attack effectiveness, evaluated on a single NVIDIA L40 GPU. MarkCleaner achieves a remarkable inference speed of 5.58 FPS (0.179 s per image), which is orders of magnitude faster than optimization-based methods (e.g., 
$400 \times$ faster than IRA, $36 \times$ faster than UnMarker). Notably, the improved efficiency does not degrade attack performance. While fast baselines often fail against the SOTA semantic watermark, our method consistently achieves near-zero detection rates. Our method breaks the trade-off between speed and attack strength, offering a real-time solution capable of defeating the SOTA semantic watermarking schemes.

\begin{table}[t]
  \centering
  \caption{Inference on a single L40 GPU. TreeRing and HSQR denote the TPR of the respective watermarking schemes. Time and FPS measure inference speed. Bold indicates the best performance.}
  \label{tab:efficiency}
  \renewcommand{\arraystretch}{1.4}
  \small
  \resizebox{1\linewidth}{!}{
    \begin{tabular}{l|cccc|c}
    \hline
    \textbf{Metric} & \textbf{IRA} & \textbf{UnMarker} & \textbf{CtrlRegen+} & \textbf{NFPA} & \textbf{Ours} \\
    \hline
    TreeRing $\downarrow$ & 0.800    & 0.010   & 0.001   & 0.003   & \textbf{0.001}   \\
    HSQR $\downarrow$      & 1.0     & 0.033   & 1.0    & 0.367   & \textbf{0.022}   \\
    \hline
    Time (s) $\downarrow$ & 72.155 & 6.527 & 2.732 & 3.887 & \textbf{0.179} \\
    FPS $\uparrow$        & 0.01    & 0.15   & 0.37   & 0.26   & \textbf{5.58}   \\
    \hline
  \end{tabular}}
\vspace{-5pt}
\end{table}

\subsection{Ablation Study}
\begin{figure}[!h]
	\centering
	\includegraphics[width=0.96\linewidth]{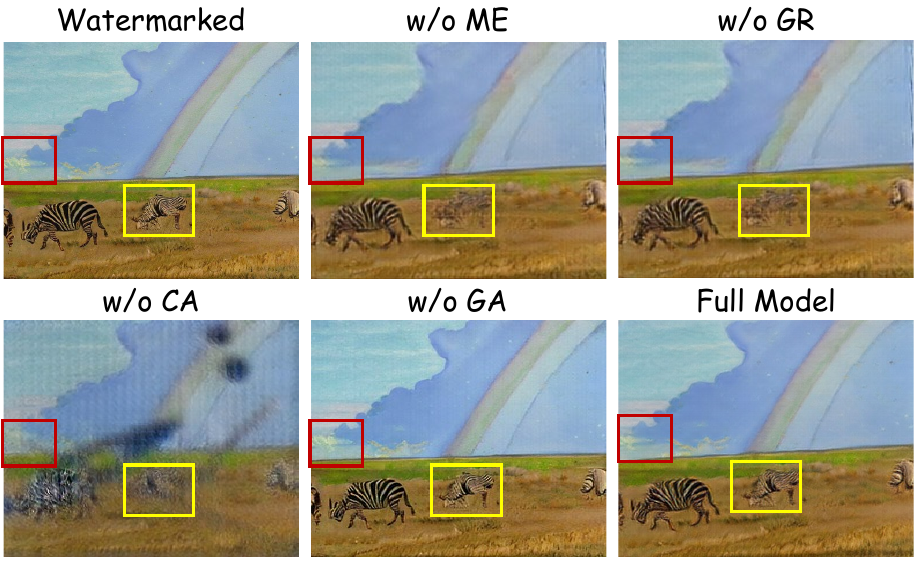}
	\caption{
    \textbf{Visualization of ablation studies.} Red boxes highlight geometric subtle shifts, while Yellow boxes highlight structural preservation. Full Model introduces imperceptible micro-shifts to erase the watermark while maintaining visual coherence.}
	\label{fig:ablation}
\end{figure}

To analyze the contribution of each component in MarkCleaner, we conduct a comprehensive ablation study by removing key modules: \textbf{Mask-Guided Encoder} (\texttt{ME}), \textbf{Gaussian Rendering} (\texttt{GR}) (replaced with a UNet decoder), \textbf{Content Alignment} (\texttt{CA}), and \textbf{Geometric Attacks supervision} (\texttt{GA}). 
The results are summarized in Figure~\ref{fig:ablation} and Table~\ref{tab:ablation}.

The results show that \texttt{GA} is the decisive factor for semantic watermark removal. Removing \texttt{GA} preserves high visual fidelity but almost completely fails to erase semantic watermarks, confirming that geometric deviation is essential for breaking the phase alignment underlying semantic watermarks. Removing \texttt{ME} leads to a significant drop in both watermark removal performance and visual quality, with particularly poor results on visible watermark removal. This indicates that explicit spatial masking is critical for accurately localizing and suppressing visible watermark regions. 
Replacing \texttt{GR} with a UNet decoder leads to reduced geometric displacement and texture blurring under misaligned supervision, reflecting that \texttt{GR} is better suited for modeling fine-grained, imperceptible geometric perturbations while preserving local details. Removing \texttt{CA} causes a severe degradation in semantic consistency, even though geometric perturbations are still applied. This confirms that semantic guidance is indispensable for maintaining high-level content structure during geometric perturbation. Overall, the full MarkCleaner framework achieves achieves near-perfect erasure ($\mathrm{TPR}\approx0$) with only a negligible trade-off in FID compared to the non-perturbed baseline, demonstrating the optimal balance between erasure and fidelity.

\begin{table}[t]
    \centering
    \caption{Ablation study of key components in MarkCleaner}
    \label{tab:ablation}
  \renewcommand{\arraystretch}{0.9}
  \resizebox{0.9\linewidth}{!}{
    \begin{tabular}{lcccc}
        \toprule
        Module & \textbf{TPR} $\downarrow$  & \textbf{FID} $\downarrow$ & \textbf{CLIP} $\uparrow$ & \textbf{LPIPS} $\downarrow$ \\
        \midrule
        w/o \texttt{ME}      & 0.2467 & 145.145 & 0.3154 & 0.6846 \\
        w/o \texttt{GR}  & 0.1367 & 146.604 & 0.3163	& 0.6837 \\
        w/o \texttt{CA}      & 0.1567 & 149.416 & 0.2528 & 0.7064 \\
        w/o \texttt{GA}  & 1.0000 & 131.487 & 0.3324 & 0.6676 \\
        \midrule
        Full Model    & \textbf{0.0001} & \textbf{133.196} & \textbf{0.3705} & \textbf{0.6295} \\
        \bottomrule
    \end{tabular}
    }
    \vspace{-6pt}
\end{table}

\section{Conclusion}

We identify that current semantic watermarks are intrinsically sensitive to geometric perturbations due to their reliance on coherent phase alignment, allowing imperceptible spatial displacements to invalidate verification without degrading image content.
Instead of aggressive content regeneration, MarkCleaner combines masked modeling for pattern-level restoration with stochastic micro-geometric displacement via 2DGS rendering, enabling watermarks removal. Extensive experiments demonstrate that MarkCleaner outperforms SOTA generative removal methods in watermark removal effectiveness, visual fidelity, and inference efficiency.
This perspective suggests that geometric robustness should be treated as a first-class consideration in future watermark design.

\clearpage
\newpage
\section*{Impact Statement}

This paper investigates the robustness of digital watermarking methods. We reveal a critical vulnerability in watermarks against micro-geometric perturbations. We acknowledge that the proposed method, Markcleaner, could potentially be misused for unauthorized watermark removal, causing copyright infringement, or evading AI-generated content detection. However, we believe that identifying and publicizing these vulnerabilities is a necessary step to a more robust and reliable watermark. By exposing the fragility of current spatial alignment mechanisms, this work motivates the development of watermarking methods that are robust against geometric distortions. We urge the community to treat geometric robustness as an important constraint in future image watermark system.

\nocite{langley00}

\bibliography{reference}
\bibliographystyle{icml2026}

\newpage
\appendix
\onecolumn

\section{Overview of Appendix.} 

The appendix provides comprehensive theoretical analysis, architectural details, and extended experimental evidence to support the main paper. We first define the threat model and scope of our attack method. We then present the formal mathematical derivation of how micro-geometric perturbations disrupt latent watermark phase coherence, and elaborate on the implementation details of the 2D Gaussian Splatting decoder and content alignment module. We include extensive qualitative and quantitative comparisons across multiple robust watermarking schemes. We also demonstrate an extension of our method to interactive visible watermark removal via user-provided masks. Finally, we critically analyze the current limitations of MarkCleaner and discuss promising directions for future research.


\section{Threat Model}
\label{subsec:threat_model}

We formulate the watermark removal problem within a realistic threat model involving two entities: a provider who operates a watermark-protected image n service, and an attacker attempting to erase watermark while retaining semantic and visual utility.

\textbf{Provider's Goals and Capabilities:} The provider deploys a generative model $\mathcal{G}$ with watermarking mechanism $\mathcal{W}$ to establish provenance and prevent misuse of generated content. Given a generated image $I$, the watermark is either embedded post-hoc or during generation. The provider controls both the generative model, watermark encoder $\mathcal{E}$, and watermark decoder $\mathcal{D}$, where $\mathcal{D}(x_w) \to \{0, 1\}$ indicates watermark presence. Detection requires only a single image and assumes the provider has sufficient computational resources to train robust watermarking systems.

\textbf{Attacker's Goals and Capabilities:} The attacker aims to produce $I' = \mathcal{T}(I)$ that removes the watermark ($\mathcal{D}(I') \to 0$) while preserving semantic and perceptual content. Watermark detection confidence drops below random chance 
($\mathrm{ACC}= 0.5$ or TPR@1\%FPR $= 0$).
The attacker operates under strict constraints:
\begin{itemize}
    \item \textbf{Access:} No knowledge of model parameters, watermarking schemes, or detector internals; no access to detection APIs for iterative queries.
    \item \textbf{Data:} No paired clean-watermarked examples; no proprietary datasets; access to only a single suspicious image.
    \item \textbf{Compute:} Limited resources excluding multi-GPU training; restricted to lightweight inference on consumer hardware.
    \item \textbf{Time:} Per-image processing must complete within seconds, recluding minutes-long optimization.
\end{itemize}

The attacker operates without knowledge of the watermark embedding algorithm, 
detection mechanism, or generative model parameters. This 
reflects realistic black-box scenarios where watermark removal must succeed 
using only the watermarked image itself, without access to the watermarking 
system or model internals.

\section{Geometric Vulnerability Analysis of Semantic Watermarks}
\label{sec:Geometry_analysis}
In this section, we provide a formal proof of how pixel-domain translations propagate through the VAE encoder to induce latent-domain phase ramps, thereby destroying watermarks embedded in the real part of the latent Fourier spectrum.

\subsection{Formalism of Real-Part Latent Watermarking}
Consider a latent diffusion model where the VAE encoder $\mathcal{E}$ maps an image $I \in \mathbb{R}^{W \times H \times 3}$ to a latent $z \in \mathbb{R}^{w \times h \times c}$, with downsampling factor $d=8$. TreeRing Watermark~\cite{Treerings_nips_2023} modifies the initial noise $z_T$, which propagates to the generated latent $z$. Let $Z_{u,v} = \mathcal{F}(z)[u,v] = A_{u,v} e^{i\phi_{u,v}}$ be the Fourier coefficient at frequency $(u,v)$. The watermark is embedded by constraining the \textbf{real part} of $Z_{u,v}$ within a frequency mask $M$ toward a target value $\eta_{u,v} \in \mathbb{R}$:
\begin{equation}
\text{Re}(Z_{u,v}) = A_{u,v} \cos(\phi_{u,v}) = \eta_{u,v}, \quad \forall (u,v) \in M.
\label{eq:embedding}
\end{equation}
Detection relies on the L1-distance: $d = \frac{1}{|M|} \sum_{M} | \text{Re}(\hat{Z}_{u,v}) - \eta_{u,v} | < \tau$, where $\hat{Z}$ is recovered via DDIM inversion.

\subsection{Phase Ramp Induction via Pixel Translation and Rotation}
We provide a rigorous derivation of how composite geometric transformations $\mathcal{T}_{\boldsymbol{\Delta}, \theta}$ (translation $\boldsymbol{\Delta}$ and rotation $\theta$) propagate through the VAE-Inversion pipeline to invalidate latent watermarks.

\textbf{Phase Modulation via Fractional Latent Shifts.}
Let $\mathcal{T}_{\boldsymbol{\Delta}}$ be a pixel-domain translation by $\boldsymbol{\Delta} = (\Delta_x, \Delta_y)$. The VAE encoder $\mathcal{E}$ can be modeled as a continuous convolution with kernels $\psi$ followed by a strided sampling operator $\mathcal{S}_d$~\cite{Aliasfree_2025_CVPR}. The latent value at discrete index $\mathbf{n}$ for the shifted image $I' = \mathcal{T}_{\boldsymbol{\Delta}}(I)$ is:
\begin{equation}
z'[\mathbf{n}] = \int_{\mathbb{R}^2} I(\mathbf{x} - \boldsymbol{\Delta}) \psi(d\mathbf{n} - \mathbf{x}) d\mathbf{x} = \int_{\mathbb{R}^2} I(\mathbf{y}) \psi(d(\mathbf{n} - \frac{\boldsymbol{\Delta}}{d}) - \mathbf{y}) d\mathbf{y}.
\end{equation}
If $\boldsymbol{\Delta}$ is a non-multiple of $d$, $\boldsymbol{\Delta}/d$ constitutes a fractional shift in the latent grid. By the Fourier Shift Theorem, this spatial shift in $z$ induces a frequency-dependent phase ramp in the latent spectrum:
\begin{equation}
\hat{Z}'_{u,v} = \mathcal{F}(z')_{u,v} = Z_{u,v} \cdot \exp\left( -i 2\pi \left( \frac{u \Delta_x}{d w} + \frac{v \Delta_y}{d h} \right) \right) + \epsilon_{\text{alias}},
\label{eq:latent_shift}
\end{equation}
where $\Delta\phi_{u,v} = -2\pi (\frac{u \Delta_x}{d w} + \frac{v \Delta_y}{d h})$ is the induced phase shift and $\epsilon_{\text{alias}}$ is the VAE aliasing noise.

\textbf{Coordinate Drift and Kernel Non-Equivariance.}
While Tree-ring employs a circular mask $M$ to mitigate rotation, its detection is \textbf{point-wise} and thus vulnerable to coordinate drift. A pixel-domain rotation $\theta$ induces a rotation of the latent frequency coordinates:
\begin{equation}
\begin{pmatrix} u' \\ v' \end{pmatrix} = \begin{pmatrix} \cos \theta & -\sin \theta \\ \sin \theta & \cos \theta \end{pmatrix} \begin{pmatrix} u \\ v \end{pmatrix}
\end{equation}
Since VAE kernels $\psi$ are fixed on a rectilinear grid, they exhibit \textbf{rotational non-equivariance}~\cite{CNNshift_2019_ICML}. The encoded features at rotated coordinates $(u', v')$ are not simply rotated versions of the original features but are distorted by the kernel's spatial bias, denoted as $\epsilon_{\text{rot}}$. This distortion, combined with the phase ramp from translation, results in a composite latent spectrum:
\begin{equation}
\hat{Z}'_{u,v} = A_{u', v'} e^{i(\phi_{u', v'} + \Delta \phi_{u,v})} + \epsilon_{\text{alias}} + \epsilon_{\text{rot}}
\label{eq:composite_latent}
\end{equation}

\subsection{Mechanism of Real-Part Signature Destruction}
Watermarks are typically embedded in the real part: $\text{Re}(Z_{u,v}) = A_{u,v} \cos(\phi_{u,v}) = \eta_{u,v}$. Under the compounded phase perturbation $\Delta \Phi'$, the recovered real part $\text{Re}(\hat{Z}'_{u,v})$ is:
\begin{equation}
\text{Re}(\hat{Z}'_{u,v}) = \underbrace{A_{u,v} \cos(\phi_{u,v})}_{\text{Signal } \eta_{u,v}} \cos(\Delta \Phi') - \underbrace{A_{u,v} \sin(\phi_{u,v})}_{\text{Imaginary Leakage}} \sin(\Delta \Phi')
\end{equation}
This \textbf{Phase-to-Real Part Leakage} leads to two failure modes:
\begin{enumerate}
    \item \textbf{Attenuation}: The original signal $\eta_{u,v}$ is suppressed by $\cos(\Delta \Phi')$. As $\Delta \Phi' \to \pi/2$, the signal vanishes.
    \item \textbf{Interference}: The high-entropy imaginary component $\text{Im}(Z)$ is projected onto the real axis via $\sin(\Delta \Phi')$, acting as incoherent noise.
\end{enumerate}

\subsection{Statistical Invalidation and Detection Failure}
Under a composite geometric transformation $\mathcal{T}_{\boldsymbol{\Delta}, \theta}$ consisting of translation $\boldsymbol{\Delta}$ and rotation $\theta$, the detection score $d'$ deviates from the original key $\eta$ due to the synergistic effects of phase scrambling and coordinate drift:
\begin{equation}
d' = \frac{1}{|M|} \sum_{(u,v) \in M} \left| A_{u_\theta, v_\theta} \cos(\phi_{u_\theta, v_\theta} + \Delta\phi_{u,v}) - \eta_{u,v} \right|,
\label{eq:composite_detection_score}
\end{equation}
where $(u_\theta, v_\theta) = \mathbf{R}_{-\theta}(u,v)$ represents the rotated frequency coordinates.

This invalidation is driven by two distinct but coupled mechanisms:

\textbf{1. Phase Ramp via Translation:} As established by the Fourier Shift Theorem, the pixel-domain translation $\boldsymbol{\Delta}$ induces a fractional shift in the latent grid due to the VAE's stride $d=8$. This manifests as a frequency-dependent phase ramp $\Delta\phi_{u,v}$. Across a ring mask $M$ with radius $r \in [r_{\min}, r_{\max}]$, the phase variation range is:
\begin{equation}
\Delta\phi_{\text{range}} \approx 4\pi \cdot \frac{r_{\max} \cdot \|\boldsymbol{\Delta}\|_2}{d \cdot w}.
\label{eq:phase_range_composite}
\end{equation}

\textbf{2. Coordinate Drift via Rotation:} Although the mask $M$ is radially symmetric, Tree-ring detection is \textbf{point-wise}. Rotation re-maps the signal originally at $(u, v)$ to a new coordinate $(u', v')$. For a frequency index at radius $r$, the spatial drift $s$ in the discrete Fourier grid is:
\begin{equation}
s \approx r \cdot \theta \quad (\text{in grid units}).
\label{eq:coord_drift}
\end{equation}

In the complex plane, these mechanisms act as a \textbf{Composite Phase-Coordinate Collapse}. Translation rotates the watermark vectors away from the real-axis key $\eta$, while rotation drifts the signal to incorrect frequency bins, effectively replacing the watermark coefficients with uncorrelated background coefficients.

\textbf{Statistical Convergence:} As the transformation magnitude increases, the distribution of the compounded phase $\Phi = \phi_{u', v'} + \Delta\phi_{u,v}$ becomes increasingly uniform over $[0, 2\pi)$. Consequently, the expectation of the recovered real part follows a Sinc-like decay:
\begin{equation}
\left| \mathbb{E}[\operatorname{Re}(\hat{Z}'_{u,v})] \right| = \left| \eta_{u,v} \cdot \frac{\sin\alpha}{\alpha} \right| \cdot \mathcal{C}(\theta),
\label{eq:expectation_decay_composite}
\end{equation}
where $\alpha = \Delta\phi_{\text{range}}/2$ and $\mathcal{C}(\theta) \in [0, 1]$ is a coherence factor representing the coordinate alignment. For Stable Diffusion ($d=8, w=64, r_{\max}=16$), we identify the following critical thresholds for reliable detection failure ($P > 0.95$):

\begin{itemize}
    \item \textbf{Translation Threshold:} $\|\boldsymbol{\Delta}\|_2 \geq 8$ pixels. At this magnitude, $\Delta\phi_{\text{range}} \geq \pi$, leading to $\alpha \geq \pi/2$ and significant signal attenuation ($\sin(\alpha)/\alpha \leq 0.637$).
    \item \textbf{Rotation Threshold:} $\theta \geq 3.6^\circ$. At this angle, the drift $s \approx 16 \cdot (3.6 \cdot \pi / 180) \approx 1.0$, causing the signal to jump to adjacent discrete frequency bins, resulting in a total mismatch with the point-wise key $\eta_{u,v}$.
    \item \textbf{Composite Effect:} A combined transformation of \textbf{8 pixels} and \textbf{$5^\circ$} ensures that $\mathbb{E}[\operatorname{Re}(\hat{Z}')] \to 0$, forcing $d'$ to converge toward the mean magnitude of the latent spectrum $\bar{A}\approx 0.886$ (for standard Gaussian noise $z_T$).
\end{itemize}

Since the embedding threshold satisfies $\tau \ll \bar{A}$, the de-synchronized vectors undergo destructive interference, causing $d' > \tau$. While specific watermark variants (e.g., constant rings) claim partial rotational invariance, our analysis underlines that the coupling of coordinate drift with VAE-induced sampling phase inconsistency ensures that even isotropic patterns suffer from coherence collapse. Furthermore, we observe that the phase-domain interference occurs at a perturbation scale significantly smaller than what would be required to manifest as perceptual magnitude erosion, confirming that phase-locked synchronization is the point of latent watermarking.

\textbf{Conclusion of Proof.}
We have proven that composite geometric transformations $\mathcal{T}_{\boldsymbol{\Delta}, \theta}$ exploit the Sampling Phase Inconsistency of the VAE. While translation scrambles the phase coherence via a linear ramp, rotation introduces coordinate misalignment and non-linear structural noise. Together, they trigger a collapse of the structured phase-locked identity required for detection, rendering the watermark statistically indistinguishable from random noise while preserving the magnitude-based visual content. 
\qed

\section{More Details of Gaussian Rendering}
\label{sec:supp_gs}

\subsection{3D Gaussian Splatting}

3D Gaussian Splatting (3DGS) \cite{3DGS_tog_2023} has emerged as a powerful technique for real-time novel view synthesis, representing 3D scenes using explicit, learnable Gaussian primitives rather than implicit neural representations. 3DGS models scenes with a large collection of 3D Gaussians combined with differentiable rasterization, enabling real-time rendering and fine-grained editability.

In 3DGS, each Gaussian primitive is parameterized by:
\begin{itemize}[]\setlength\itemsep{-0em}
    \vspace{-6pt}
    \item \textbf{Position} $\mu \in \mathbb{R}^3$: the 3D center of the Gaussian,
    \item \textbf{Covariance} $\Sigma \in \mathbb{R}^{3\times 3}$: a 3D covariance matrix describing the shape and orientation,
    \item \textbf{Opacity} $\alpha \in \mathbb{R}$: controls the transparency,
    \item \textbf{Color}: represented via spherical harmonics (SH) coefficients for view-dependent appearance.
    \vspace{-6pt}
\end{itemize}

The 3D covariance matrix $\Sigma$ is typically decomposed into a rotation matrix $R$ and a scaling matrix $S$:
\begin{equation}
    \Sigma = RSS^\top R^\top
\end{equation}
This parameterization ensures positive semi-definiteness while providing intuitive control over the Gaussian's shape.

For rendering, 3D Gaussians are projected onto the 2D image plane. Given a viewing transformation $W$ and the Jacobian $J$ of the affine approximation of the projective transformation, the projected 2D covariance is:
\begin{equation}
    \Sigma' = JW\Sigma W^\top J^\top
\end{equation}

The final pixel color is computed via alpha blending of overlapping Gaussians sorted by depth:
\begin{equation}
    C = \sum_{i \in \mathcal{N}} c_i \alpha_i \prod_{j=1}^{i-1}(1-\alpha_j)
\end{equation}
where $c_i$ is the color and $\alpha_i$ is the opacity multiplied by the Gaussian's contribution at that pixel.

\subsection{2D Gaussian Splatting}

Building upon 3DGS, 2D Gaussian Splatting adapts similar principles for 2D image processing. The image is represented by numerous 2D Gaussians, each characterized by mean $\mu_i \in \mathbb{R}^2$, covariance matrix $\Sigma_i \in \mathbb{R}^{2\times 2}$, opacity $\alpha_i \in \mathbb{R}$, and color $c_i \in \mathbb{R}^3$.

To ensure positive semi-definiteness, the covariance is parameterized as $\Sigma_i = L_i L_i^\top$, where $L_i \in \mathbb{R}^{2\times 2}$ is a lower triangular matrix:
\begin{equation}
    L_i = \begin{pmatrix} l_{11} & 0 \\ l_{21} & l_{22} \end{pmatrix}
\end{equation}
This reduces the covariance representation from 4 to 3 parameters. We denote the complete parameter set as the Gaussian space $\Theta = \{\mu, L, c, \alpha\}$.

The pixel value at position $\mathbf{p} = [x,y]^\top$ is computed as:
\begin{equation}
    I_{\mathbf{p}} = \sum_i c_i \alpha_i \exp\left(-\frac{1}{2}(\mathbf{p} - \mu_i)^\top \Sigma_i^{-1} (\mathbf{p} - \mu_i)\right)
\end{equation}
This formulation leverages the favorable mathematical properties of Gaussian Mixture Models, where complex distributions can be constructed by multiple Gaussian kernels.

\subsection{More Details of Gaussian Modeling}
\label{sec:GS}

Given a masked input image $I_{\text{mask}} \in \mathbb{R}^{3\times H\times W}$, we employ a UNet encoder to extract Gaussian features. The encoder outputs a dense feature map $F_g \in \mathbb{R}^{C'\times H\times W}$, which is downsampled via strided convolution to Gaussian-level features $F_g' \in \mathbb{R}^{N\times C'}$, where $N$ denotes the number of Gaussian kernels.

The encoder's hierarchical structure with skip connections captures rich context while enhancing training stability. The Gaussian mean $\mu$, empirically observed to be sensitive during training, is initialized uniformly across the 2D plane as $\mu_{fix}$, and only the offset $\mu_{bias}$ is learned.

Specifically, Gaussian parameters are decoded from $F_g'$ using lightweight MLPs:
\begin{equation}
    \mu_{bias} = E_\mu(F_g'), \quad c = E_c(F_g'), \quad \sigma = E_\sigma(F_g'), \quad \alpha = E_\alpha(F_g')
\end{equation}
where $\mu_{bias} \in \mathbb{R}^{N\times 2}$ denotes learnable positional offsets, $c \in \mathbb{R}^{N\times 3}$ represents color, $\sigma \in \mathbb{R}^{N\times 3}$ represents covariance parameters, and $\alpha \in \mathbb{R}^{N\times 1}$ is the opacity.

To constrain the spatial extent of $\mu_{bias}$, we apply $\texttt{tanh}$ activation, restricting values to $[-1, 1]$. Final positions are obtained by: $\mu = \mu_{bias} + \mu_{fix}$.

These parameters are subsequently rasterized onto the image plane, and their soft overlapping nature guarantees smooth color and intensity transitions across pixels.

\subsection{Patch-Level Rasterization}

High-fidelity image representation requires numerous Gaussians, leading to significant GPU memory consumption. Following the tile-based rasterization of 3DGS\cite{3DGS_tog_2023}, we adpot patch-level rasterization to address this challenge.

An image $I \in \mathbb{R}^{H\times W}$ is divided into $N_p = \frac{H}{p} \times \frac{W}{p}$ non-overlapping patches of size $p \times p$. Each patch $(i,j)$ maintains a dedicated set of $N_{patch}$ Gaussians. The parameter space transforms into:
\begin{equation}
    \mu \in \mathbb{R}^{N_p\times N_{patch}\times 2}, \quad c \in \mathbb{R}^{N_p\times N_{patch}\times 3}, \quad l \in \mathbb{R}^{N_p\times N_{patch}\times 3}
\end{equation}

While this may not reduce the total number of Gaussians ($N_p \times N_{patch}$), it reduces concurrent memory demand during rasterization and enables parallel rendering across patches.

\textbf{Overlap Blending.} To avoid boundary discontinuities, each patch is rendered over an extended $(p+2a) \times (p+2a)$ region including $a$-pixel borders on all sides. The final image retains central $(p-2a) \times (p-2a)$ regions and blends overlapping borders using distance-weighted averaging, ensuring smooth transitions.

\section{Content Alignment Details}
\label{sec:supp_dino}

To ensure global semantic consistency during reconstruction, we incorporate pretrained DINOv2 features as high-level guidance. This is particularly important when using patch-level processing, as it helps maintain coherent semantics across patch boundaries.

\subsection{DINOv2 Feature Extraction}

DINOv2 is a self-supervised vision transformer that learns robust visual features without labels. Its features capture high-level semantic structure and exhibit robustness to appearance variations\cite{DINOV2_tmlr_2024}.

We extract features from a pretrained DINOv2 ViT model (frozen during training). However, when the input image is heavily masked, the reliability of DINOv2 features deteriorates. Inspired from \cite{2dgsinpainting_2025}, the features extracted from lightly masked inputs still preserve rich semantic information, whereas increasing mask ratios lead to progressively degraded representations. Therefore, we employ a lightweight MLP to transform masked features into estimated clean features:
\begin{equation}
    f_{pred} = \text{MLP}(f_{masked})
\end{equation}
where $f_{\text{masked}}$ denotes DINOv2 features extracted from the masked input image, and $f_{\text{pred}}$ is the adapted semantic representation.

\subsection{Adaptive Layer Normalization}
\begin{figure}[h]
    \centering
    \includegraphics[width=0.9\linewidth]{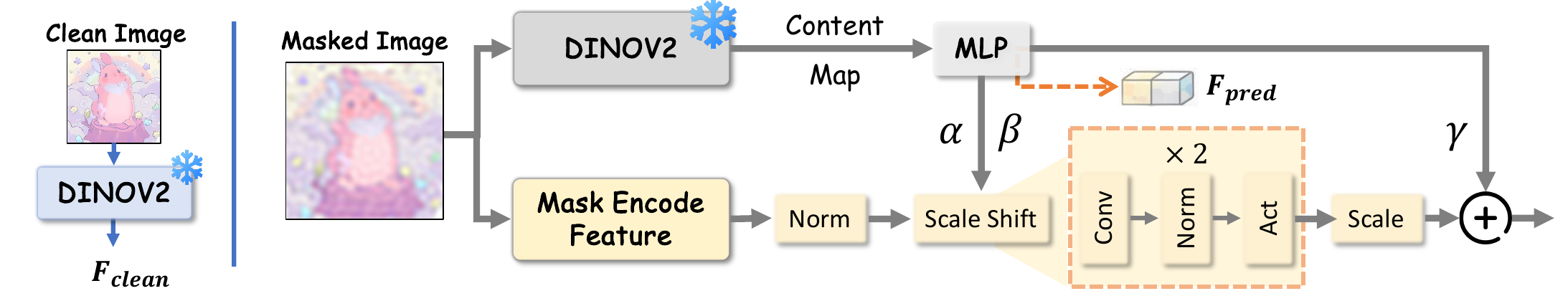}
    \caption{\textbf{Content Alignment Block.} DINO features are processed by an MLP to predict scale ($\alpha, \beta$) and shift ($\gamma$) parameters for Adaptive Layer Normalization, conditioning the decoder on semantic context.}
    \label{fig:supp_adaln}
\end{figure}

We inject the semantic features into the decoder using Adaptive Layer Normalization (AdaLN) \cite{analn_2019_cvpr}, as illustrated in Fig.~\ref{fig:supp_adaln}. Given a predicted semantic feature vector $f_{pred} \in \mathbb{R}^{C}$ and a hidden feature map $h \in \mathbb{R}^{C \times H \times W}$, the AdaLN operation is defined as:
\begin{equation}
    \text{AdaLN}(h, f_{pred}) = B(\text{LN}(h) \cdot \alpha + \beta) \cdot \gamma
\end{equation}
where $\text{LN}(\cdot)$ denotes Layer Normalization, $B(\cdot)$ denotes the main convolutional block, and $\alpha, \beta, \gamma \in \mathbb{R}^{C}$ are scale, shift, and output modulation parameters predicted from $f_{pred}$ via a lightweight MLP. This modulation is applied at multiple decoder layers, allowing semantic information to guide reconstruction at different scales.

\section{Loss Function Details}
\label{sec:supp_loss}
Here, $I_{out}$ denotes the rendered output image produced by the Gaussian decoder, 
and $I_{pert}$ denotes the geometrically perturbed target image used for supervision.
The training objective is formulated as a weighted combination of four loss terms:
\begin{equation}
    \mathcal{L}_{total} = \lambda_1 \mathcal{L}_{rec} + \lambda_2 \mathcal{L}_{P} + \lambda_3 \mathcal{L}_{G} + \lambda_4 \mathcal{L}_{D},
\end{equation}
where the weights are empirically set as $\lambda_1 : \lambda_2 : \lambda_3 : \lambda_4 = 1 : 0.3 : 3 : 1$.

\textbf{Reconstruction Loss.} 
The reconstruction loss measures the difference between the rendered output $I_{out}$ and the perturbed target $I_{pert}$ in both spatial and frequency domains:
\begin{equation}
    \mathcal{L}_{rec} = \mathcal{L}_{1} + \lambda_{freq} \mathcal{L}_{freq},
\end{equation}
where $\mathcal{L}_{1} = \|I_{out} - I_{pert}\|_1$ is the spatial L1 loss, and $\lambda_{freq} = 1$.

For the frequency loss, we define a frequency distance metric that considers both amplitude and phase. Regarding each frequency value $\mathcal{F}(I)(u,v)$ as a two-dimensional Euclidean vector $\vec{f}$, the magnitude corresponds to amplitude while the angle corresponds to phase. We define the frequency distance $\mathcal{D}(\cdot,\cdot)$ as \cite{MFM_iclr_2023}:
\begin{equation}
\begin{aligned}
\mathcal{D}(\vec{f}_{out},\vec{f}_{pert}) &= \|\vec{f}_{out}-\vec{f}_{pert}\|_{2}^{\gamma} = |\mathcal{F}(I_{out})(u,v)-\mathcal{F}(I_{pert})(u,v)|^{\gamma} \\
&= \left((\mathcal{R}_{out}(u,v)-\mathcal{R}_{pert}(u,v))^{2}+(\mathcal{I}_{out}(u,v)-\mathcal{I}_{pert}(u,v))^{2}\right)^{\gamma/2}.
\end{aligned}
\end{equation}
where $\mathcal{R}$ and $\mathcal{I}$ are the real and imaginary parts of the Fourier transform, and $\gamma$ controls the sharpness of the distance function (set to 1 by default). The frequency loss is the average distance over all spectrum positions:
\begin{equation}
    \mathcal{L}_{freq} = \frac{1}{HW}\sum_{u=0}^{H-1}\sum_{v=0}^{W-1}|\mathcal{F}(I_{out})(u,v)-\mathcal{F}(I_{pert})(u,v)|^{\gamma}.
\end{equation}

\textbf{Perceptual Loss.} We use LPIPS~\cite{lpips_cvpr_2018} to capture perceptual similarity:
\begin{equation}
    \mathcal{L}_{P} = \text{LPIPS}(I_{out}, I_{pert}).
\end{equation}

\textbf{Adversarial Loss.} We employ a discriminator $D$ with hinge loss:
\begin{align}
    \mathcal{L}_{G}^{D} &= \mathbb{E}[\max(0, 1-D(I_{pert}))] + \mathbb{E}[\max(0, 1+D(I_{out}))] \\
    \mathcal{L}_{G}^{G} &= -\mathbb{E}[D(I_{out})].
\end{align}

\textbf{Content Alignment Loss.} To explicitly align the predicted features with the ground truth semantic representations, we use a feature alignment loss based on cosine similarity:
\begin{equation}
    \mathcal{L}_{D} = 1 - \frac{f_{pred} \cdot f_{clean}}{\|f_{pred}\|_2 \cdot \|f_{clean}\|_2}.
\end{equation}
where $f_{pred}$ is the semantic feature predicted by the MLP from the masked input (as described in Sec.~\ref{sec:supp_dino}), and $f_{clean}$ is the DINOv2 feature extracted from the original input image. This loss encourages the predicted features to align with the clean features, enhancing global semantic consistency.

\section{More Experimental Results}
\subsection{Image Watermark Removal settings}\label{Attack_settings}

We evaluate MarkCleaner against a diverse set of watermark removal methods categorized into three groups.

\textbf{Distortion Attacks} comprise standard signal perturbations commonly adopted in prior watermark removal studies, including JPEG compression (quality factor 20), cropping and scaling (scale factor 0.8), Gaussian blur (kernel size 15), Gaussian noise (standard deviation 0.3), color jitter (brightness factor 4), rotation (15 degrees), and translation (20 pixels). Each perturbation is applied individually to assess robustness against basic image transformations.

\textbf{Regeneration Attacks} leverage generative priors to reconstruct images. 
Specifically, we consider Diffusion-Attack(DA)~\cite{diffuction-attack_iclr_2024} using DDIM inversion with 50 inference steps and guidance scale 7.5; VAE-Attack (VA-B) based on variational image compression with a scale hyperprior~\cite{VAE_B_iclr_2018}, which reconstructs images through a learnedrate-distortion bottleneck at quality level 4: VAE-Attack (VA-C)~\cite{VAEattack_CVPR_2020} based on the Stable Diffusion VAE encoder-decoder pipeline at quality level 3; CtrlRegen+~\cite{CtrlRegen_iclr_2025} with clean-noise initialization, employing ControlNet~\cite{Controlnet_iccv_2023} with 50 denoising steps and conditioning scale 1.0 and guidance scale of 2.0, performing image regeneration via a latent diffusion process with a UniPC multistep scheduler~\cite{unipc_NIPS_2023}; and NFPA~\cite{NFPA_nips_2025} with noise perturbation strength 0.1 and 25 purification steps, using a DDIM scheduler with 50 inference steps.

\textbf{Optimization-based Attack} is represented by UnMarker~\cite{UnMarker_ISSP_2025}, a universal framework for heterogeneous watermarks, which applies a center-crop preprocessing (crop ratio 0.9) followed by a low-frequency-constrained optimization stage, configured with an initial learning rate of 0.01 under an adaptive Adam optimizer, 10 optimization iterations, and using VGG-based perceptual loss to preserve visual fidelity. We exclude surrogate-dependent method IRA ~\cite{semantic_forgery_CVPR_2025} with 100 steps optimization to ensure a strictly black-box evaluation setting.

\subsection{Additional Results of Visual Recovery}
\label{sec:sup_visual_results}
To further analyze the balance between watermark detectability and visual quality, we provide extensive qualitative comparisons to demonstrate the visual quality of MarkCleaner across different watermarking schemes.

\begin{figure}[!h]
    \centering
        \subfloat{
        \includegraphics[width=1\columnwidth]{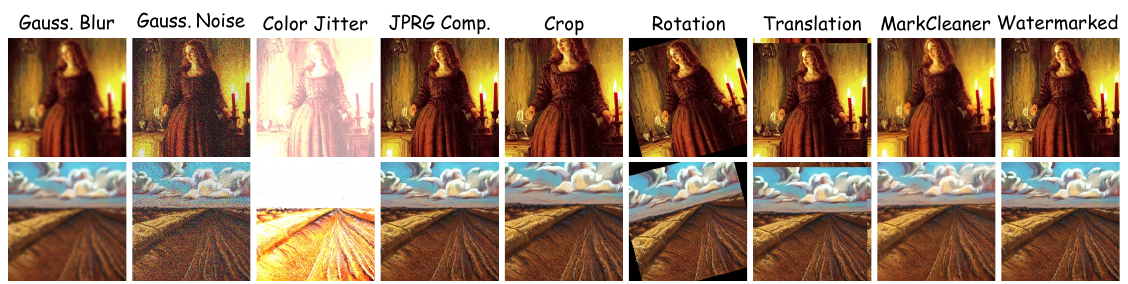}
        \label{fig:sup_treering_a}
    } \par
        \subfloat{
        \includegraphics[width=1\columnwidth]{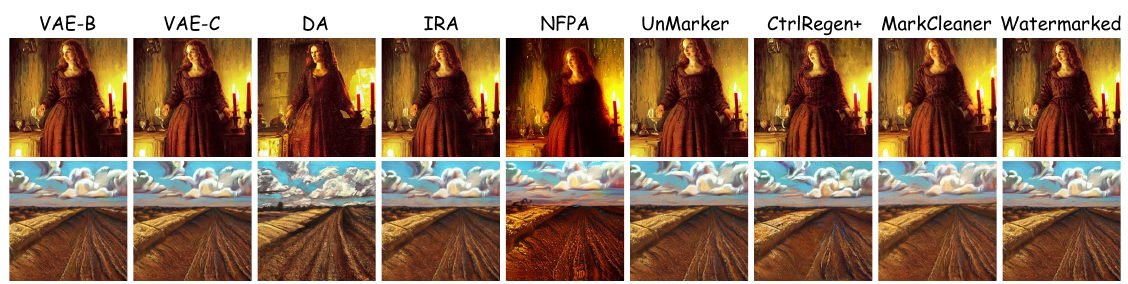}
        \label{fig:sup_treering_b}
    }
    \caption{More qualitative comparisons of watermark removal methods on images embedded with Tree-Ring watermarks \cite{Treerings_nips_2023}.}
    \label{fig:sup_treering}
\end{figure}

\begin{figure}[!h]
  \centering
  \subfloat{
    \includegraphics[width=1\columnwidth]{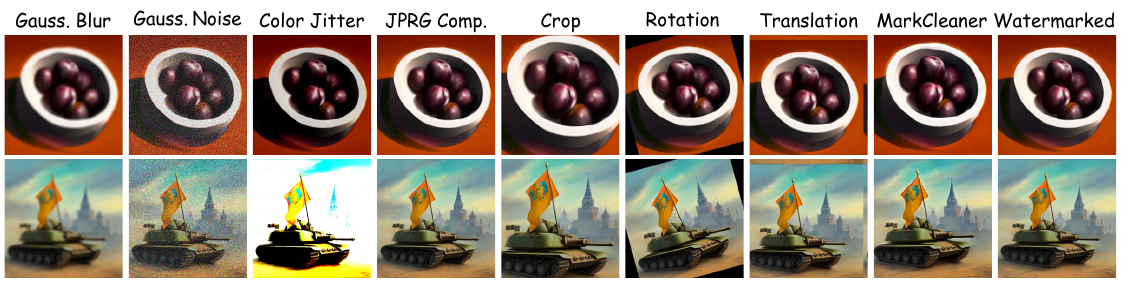}
    \label{fig:sup_ringid_a}
  } \par
  \subfloat{
    \includegraphics[width=1\columnwidth]{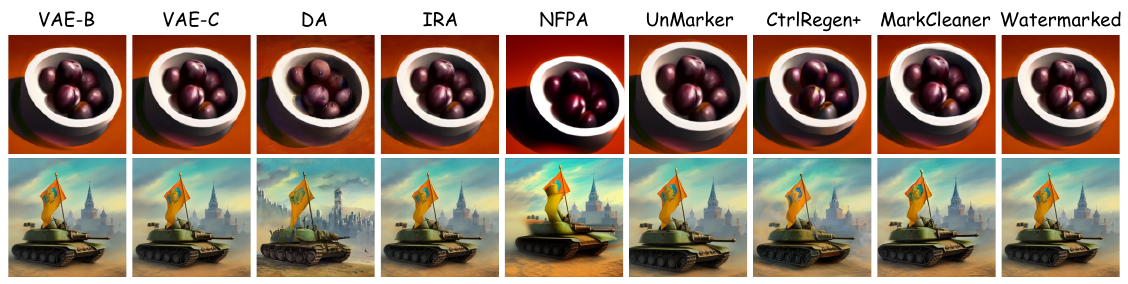}
    \label{fig:sup_ringid_b}
  }
  \caption{More qualitative comparisons of watermark removal methods on images embedded with RingID watermarks \cite{RingID_eccv_2025}.}
    \label{fig:sup_vis_ringid}
\end{figure}

\begin{figure}[t]
  \centering
  \subfloat{
    \includegraphics[width=1\columnwidth]{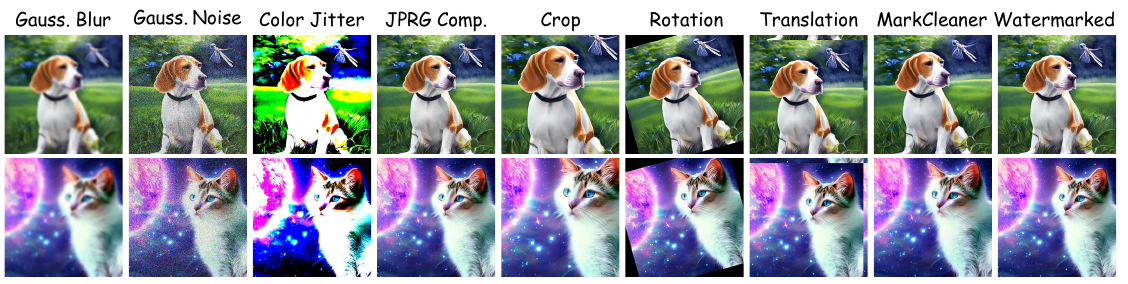}
    \label{fig:sup_HSTR_a}
  } \par
  \subfloat{
    \includegraphics[width=1\columnwidth]{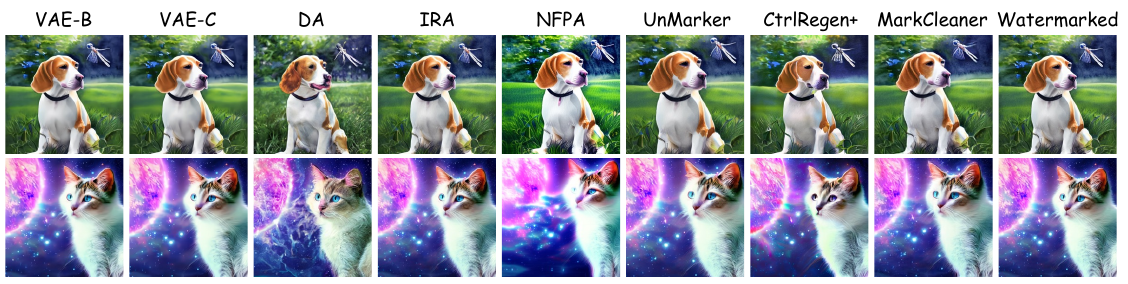}
    \label{fig:sup_HSTR_b}
  }
  \caption{More qualitative comparisons of watermark removal methods on images embedded with HSTR \cite{semantic_iccv_2025}.}
    \label{fig:sup_vis_HSTR}
\end{figure}

\begin{figure}[!h]
  \centering
  \subfloat{
    \includegraphics[width=1\columnwidth]{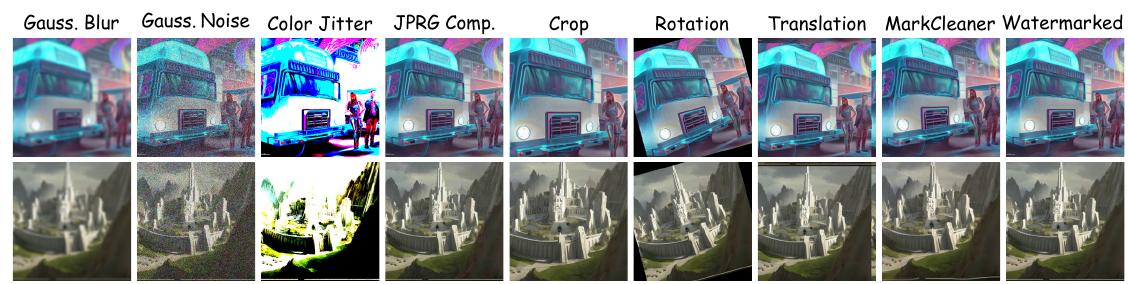}
    \label{fig:sup_HSQR_a}
  } \par
  \subfloat{
    \includegraphics[width=1\columnwidth]{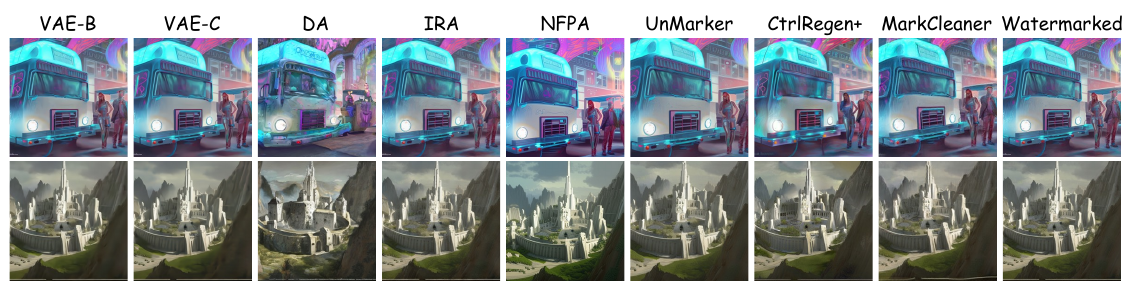}
    \label{fig:sup_HSQR_b}
  }
  \caption{More qualitative comparisons of watermark removal methods on images embedded with HSQR \cite{semantic_iccv_2025}.}
    \label{fig:sup_vis_HSQR}
\end{figure}

\textbf{Additional Qualitative Results}
Figures \ref{fig:sup_treering},  \ref{fig:sup_vis_ringid}, \ref{fig:sup_vis_HSTR} and \ref{fig:sup_vis_HSQR} present detailed visual comparisons on semantic watermark with TreeRing \cite{Treerings_nips_2023}, RingID \cite{RingID_eccv_2025} HSTR \cite{semantic_iccv_2025} and HSQR \cite{semantic_iccv_2025}, respectively, which represent four challenging semantic watermarking methods with strong robustness against conventional attacks. These results demonstrate that MarkCleaner consistently achieves superior watermark removal while preserving perceptual fidelity across diverse semantic watermarking schemes, without introducing the semantic drift, color distortion, or structural artifacts characteristic of baseline methods.

\textbf{The Discrepancy between Pixel and Perceptual Metrics.} As shown in Table \ref{tab:sup_visual_quant}, we observe a significant discrepancy in the evaluation of MarkCleaner. While perceptual metrics like FID (133.6) and LPIPS (0.627) indicate high visual fidelity comparable to the original image, standard pixel-wise metrics (Raw PSNR: 16.52, SSIM: 0.444) appear unexpectedly low. This is an inherent characteristic of our \textit{micro-geometric perturbation} strategy. Standard PSNR assumes perfect spatial alignment; however, our method deliberately introduces imperceptible spatial shifts to break the watermark's phase coherence. A shift of merely 2-3 pixels can drastically penalize PSNR without degrading the actual visual content.

To decouple geometric displacement from actual content degradation, we compute \textbf{Aligned-PSNR} and \textbf{Aligned-SSIM} \cite{alignpsnr_2021_CVPR}.
These metrics first register the output to the input using optical flow-based alignment to compensate for spatial shifts, then compute PSNR/SSIM on the aligned pair. This isolates content fidelity from geometric misalignment.
The results reveal a striking pattern: MarkCleaner's PSNR improves dramatically from 16.52 (Raw) to \textbf{25.30} (Aligned), and SSIM jumps from 0.444 to \textbf{0.820}. This substantial recovery ($>8$ dB gain) confirms that the low raw scores are primarily driven by spatial misalignment rather than semantic corruption. In contrast, generative methods like DA and IRA show minimal improvement after alignment, indicating their low fidelity stems from irreversible content hallucinations that cannot be corrected by geometric registration. The combination of high Aligned-PSNR and near-perfect watermark suppression (as shown in Tabel \ref{tab:all_results}) validates our core thesis: MarkCleaner successfully isolates geometric perturbations to disrupt watermark detection while preserving semantic integrity.

\begin{table}[htbp]
  \centering
  \caption{Quantitative comparison of visual quality metrics across different watermark removal methods}
    \begin{tabular}{c|c|ccccccc|c}
    \toprule
    Metrics & Watermarked & VAE-B & VAE-C & DA    & CtrlRegen+ & UnMarker & NFPA  & IRA & Ours \\
    \midrule
    \textbf{PSNR} $\uparrow$  & -- & 32.72 & 32.62 & 18.92 & 21.47 & 14.85 & 29.06 & 12.76 & 16.52 \\
    \textbf{SSIM} $\uparrow$ & -- &  0.893 & 0.885 & 0.511 & 0.660 & 0.434 & 0.845 & 0.358 & 0.444 \\
    \textbf{AlignPSNR} $\uparrow$ & -- & 30.54 & 30.47 & 18.39 & 28.52 & 28.96 & 21.43 & 19.17 & 25.30 \\
    \textbf{AlignSSIM} $\uparrow$ & -- & 0.924 & 0.918 & 0.495 & 0.886 & 0.917 & 0.682 & 0.666 & 0.820 \\
    \textbf{LPIPS} $\downarrow$ & 0.620 & 0.614 & 0.615 & 0.644 & 0.618 & 0.614 & 0.609 & 0.641& 0.627 \\
    \textbf{FID} $\downarrow$  &131.5 & 142.5 & 145.7 & 133.8 & 134.0 & 130.9 & 142.2 & 135.04 & 133.6 \\
    \textbf{CLIP} $\uparrow$ & 0.380 &0.386 & 0.385 & 0.356 & 0.382 & 0.386 & 0.361 & 0.3593 & 0.376 \\
    \bottomrule
    \end{tabular}%
  \label{tab:sup_visual_quant}%
\end{table}%

\subsection{Potential Application: Removal of Visible Watermarks}
Although MarkCleaner is primarily designed for robust invisible watermarks, its architecture naturally supports visible watermark removal. Our framework employs a mask-guided encoder that reconstructs image content from partial observations.
Theoretically, if the mask specifically targets regions occluded by visible watermarks, the model should be able to hallucinate and restore the underlying content.

\begin{figure}[!h]
    \centering
    \includegraphics[width=1\linewidth]{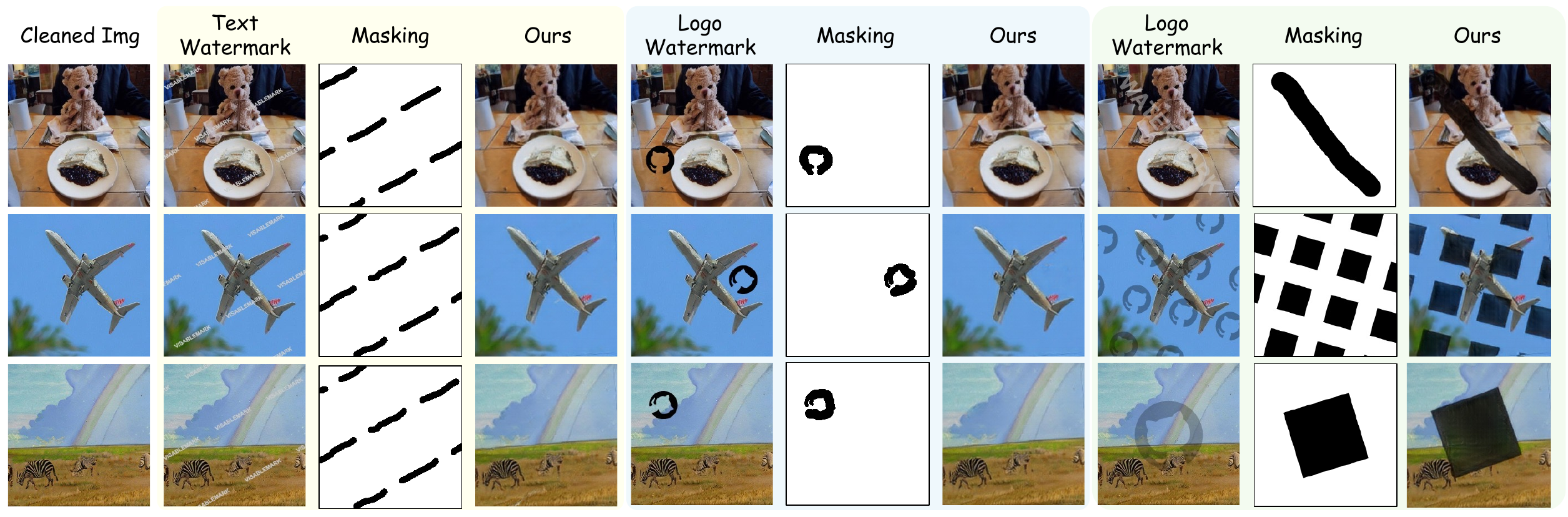}
    \caption{Application of MarkCleaner for interactive visible watermark removal.}
    \label{fig:supp_visable}
    \vspace{-5pt}
\end{figure}

To verify this capability, we conduct an exploratory experiment in which the watermark regions are manually masked at inference time. As shown in Fig.~\ref{fig:supp_visable}, MarkCleaner is able to effectively remove visible watermarks and reconstruct plausible content in the masked areas. This setting simulates an interactive scenario in which users explicitly select visible watermark regions for removal.

However, we observe a performance limitation when dealing with large, contiguous watermark patterns. Since MarkCleaner is trained using a \textit{random discrete masking strategy} (e.g., localized random drops) to learn robust features, it is less optimized for inpainting large, continuous holes often required for visible watermark removal. This domain gap between training (discrete masks) and inference (continuous block masks) can lead to blurriness or artifacts in the restored regions. We identify this as a direction for future work, where incorporating block-wise masking strategies during training could further enhance performance on visible watermarks.

\section{Discussion and Future Work}
\label{sec:discussion}
\textbf{Discussion.} While MarkCleaner effectively removes almost invisible watermarks, we observe two limitations. Firstly, our ablation studies underscore that geometric augmentation (GA) is indispensable for effective erasure; while disabling GA yields superior reconstruction fidelity, it fails to invalidate watermark detection. This binary outcome confirms our hypothesis that phase displacement, rather than semantic alteration, is the primary driver of removal. Critically, the high visual quality observed in the absence of GA defines the 'fidelity ceiling' for this paradigm. This suggests that while our current implementation achieves effective removal, there is a significant opportunity to further refine the micro-perturbations to bridge this fidelity gap, pointing towards a future where phase scrambling becomes entirely imperceptible. Second, performance degrades when handling large continuous visible watermarks. This stems from a training-inference mismatch: our random discrete masking strategy is optimized for scattered perturbations rather than block-wise inpainting. Future work could address this by incorporating continuous mask patterns during training.

\textbf{Future Work.} A compelling extension of this work lies in closing the loop between watermark removal and robust watermark design. Future research should incorporate differentiable geometric operators into adversarial training to enhance watermark robustness. By regularizing encoders against the phase-scrambling predicted by the Fourier Shift Theorem, we can develop geometry-invariant provenance. This approach targets signatures that remain detectable even under the sampling artifacts of non-equivariant, strided architectures.

\end{document}